\documentclass[english,tightenlines,eqsecnum,amsmath,amssymb,aps,prd,superscriptaddress,nofootinbib,showpacs,floats]{article}
\usepackage{amsmath,amsfonts,amssymb,multicol}
\usepackage{graphicx,caption,subcaption}
\usepackage{epstopdf}
\usepackage{cite}
\usepackage{enumitem}
\usepackage[usenames]{xcolor}
\usepackage{epstopdf}
\definecolor{AHZ}{rgb}{0.0,1,0.0}
\usepackage[colorlinks,linkcolor=AHZ,citecolor=red,bookmarks]{hyperref}
\usepackage{rotating}
\usepackage[mathscr]{eucal}
\usepackage{graphicx}
\usepackage[T1]{fontenc}
\usepackage{babel}
\usepackage{authblk}
\usepackage{epsfig}
\usepackage{color}
\usepackage{amssymb}
\usepackage{fancyhdr}

\def\nn{\nonumber\\}
\newcommand{\f}[2]{\frac{#1}{#2}}
\def\be{\begin{equation}}
\def\ee{\end{equation}}
\def\bea{\begin{eqnarray}}
\def\eea{\end{eqnarray}}

\def\bwt{\begin{widetext}}
	\def\ewt{\end{widetext}}

\usepackage{amsmath}

\usepackage{amssymb}

\begin{document}
	
\title{Novel Brans-Dicke Wormhole Solutions in Non-Vacuum Spacetimes}
\author{Amir Hadi Ziaie\thanks{ah.ziaie@riaam.ac.ir}~}
\author{Naser Sadeghnezhad\thanks{nsadegh@riaam.ac.ir}~~}
\author{~Akbar Jahan\thanks{jahan@riaam.ac.ir}}
\affil{{\rm Research~Institute~for~Astronomy~and~Astrophysics~of~ Maragha~(RIAAM), University of Maragheh,  P.~O.~Box~55136-553,~Maragheh, Iran}}
\renewcommand\Authands{and}
\maketitle
\begin{abstract}
In the present study we search for a new class of wormhole solutions in the framework of Brans-Dicke (BD) theory in the presence of anisotropic matter distribution. Considering a linear equation of state (EoS) between radial pressure and energy density profile we find exact static spherically symmetric solutions to the BD field equations which represent wormhole configurations. The solutions we obtain include both cases with zero and nonzero redshift functions, for which, the conditions on wormhole geometry together with the weak (WEC) and null (NEC) energy conditions put constraints on model parameters such as, the BD coupling and EoS parameters. These constraints also depend on other model parameters such as, the value of BD scalar field and energy density at the wormhole throat. The regularity of the obtained solutions is verified by calculating the Kretschmann scalar in order to ensure that curvature singularities are absent in the wormhole spacetime. We then find that BD wormholes in the presence of anisotropic matter can exist without violating NEC and WEC, either at the throat or across the entire spacetime. 
\end{abstract}
\maketitle 
\section{Introduction}
The field equation of general relativity (GR), being local and frame independent in nature, admits surprising solutions with non-trivial topology, among which, wormholes have been extensively investigated in the literature~\cite{FLoboBook,khu}. These objects are topological handles in spacetime connecting
widely separated parts of a single Universe or tunnel-like structures joining two different spacetimes. The study of wormholes in GR was initially triggered by the work of Ludwig Flamm in 1916 who tried to describe the Schwarzschild solution utilizing 2D embedding diagram of the Schwarzschild wormhole~\cite{Flamm}. In 1935, Einstein and Rosen constructed wormhole type solutions considering
an elementary particle model as a bridge connecting two identical sheets. This mathematical representation of space being
connected by a wormhole type solution is popularly known as the Einstein-Rosen bridge~\cite{ERose}. After this work, the interest in wormhole structures made its progress slowly over the years, through the works of Misner and Wheeler~\cite{misner-wheeler,Wheelerworm} who constructed hypothesized \lq{}\lq{}{geon}\rq{}\rq{} solutions, Graves and Brill who investigated wormhole-type solutions in Reissner-Nordstr\"{o}m spacetime~\cite{GrBrRN}, Homer Ellis who introduced the concept of drainhole~\cite{Homers} and Bronnikov’s tunnel-like solutions~\cite{Bronnsols}, see also~\cite{hisworm} for recent reviews. However, It was only in 1988 that a profound and fundamental progress in wormhole physics took place through the pioneering papers by Morris, Thorne and Yurtsever~\cite{mt,mt1}. They constructed exact static spherically symmetric solutions to GR field equation which represent Lorentzian traversable wormholes. Such amazing structures which are widely known as Morris-Thorne (MT) wormholes allow a round trip between two parts of the spacetime through the wormhole throat. For an arbitrary static spacetime one may define the wormhole throat as a two-dimensional constant-time hypersurface of minimal area, for which, the trace of extrinsic curvature vanishes and the flare-out condition holds~\cite{VissHoch1997}. As stated by Morris and Thorne~\cite{mt}, the fundamental flare-out condition can only be satisfied by violating the NEC, i.e., ${\rm T}_{\mu\nu}k^\mu k^\nu\geq0$ for all null vectors $k^\mu$, where ${\rm T}_{\mu\nu}$ is the energy-momentum tensor (EMT) of matter distribution. Matter fields that violate the NEC are called exotic~\cite{khu,khu1} and examples of such exotics fields that are known to occur in nature are: Squeezed quantum vacuum states including electromagnetic and other (non-Maxwellian) quantum fields~\cite{mt},\cite{Drumm2004}, Gravitationally squeezed vacuum electromagnetic zero-point fluctuations~\cite{Gsqueeze}, Casimir Effect, i.e., the Casimir vacuum in flat or curved space~\cite{qftcasm} and other quantum fields, states or effects~\cite{edavis2012}. In this regard, many efforts have been devoted to build and study wormhole structures in the presence of exotic matter~\cite{exo0,exo1,exo2,phantworm,intdarksec}. More proposals have addressed the possibility of minimizing the exoticity of material threading the wormhole throat, e.g., utilizing the cut and paste method~\cite{cutpaste} and thin-shell Lorentzian wormholes~\cite{exo1},\cite{thinshells}, see also~\cite{minexot},\cite{lobocgqreview} and references therein.
\par 
In modified theories of gravity, extra geometrical attributes (not present in GR) are speculated to be the source of such exotic matter~\cite{modsgrex}. In this connection, wormhole configurations have been studied in different settings in the framework of these theories, for example, traversable wormholes in higher-dimensional GR~\cite{higherdimw}, Kaluza-Klein gravity~\cite{KKGworm}, Einstein-Gauss-Bonnet gravity~\cite{gmfl}, brane world scenarios~\cite{braneworm}, Lovelock~\cite{LOVEWORM} and $f({\rm R})$ gravity theories~\cite{fr},\cite{fr1}, {massive gravity~\cite{massiveworm},} Einstein-Cartan theory~\cite{ecworm}, modified gravities with curvature-matter coupling~\cite{Garcia-Lobo}, Rastall gravity~\cite{rastallworm} and other theories~\cite{otherworms}, {see also~\cite{newast2024} for a comprehensive study of wormhole spacetimes in different modified gravity theories.} The BD theory~\cite{BDTH} has also benefited from a careful study of wormhole structures which was initiated by the works of Ellis and Bronnikov~\cite{Homers},\cite{Bronnsols}. The authors of~\cite{bd} found a static spherically symmetric vacuum solution for which if the coupling parameter satisfies the range $\omega<-2$, the BD scalar field can act as the source of exotic matter providing thus a two-way traversable wormhole. {Traversable wormhole configurations extracted from three of the Brans solutions~\cite{brans1962} have been reported by Nandi et al.~\cite{bd1} for $\omega<-2$ and $0<\omega<\infty$}. In~\cite{bd2} two new classes of exact solutions in vacuum BD theory were obtained, each of which represent a two-way
traversable wormhole for $\omega<-2$ or $-2<\omega\leq0$. Analytical non-vacuum wormhole solutions in BD theory were introduced in~\cite{bd4}, where, the authors figured out that there exists some regions of the space-parameter in which the BD {scalar field may imitate the properties of exotic matter, implying that there is a possibility to establish wormhole-like spacetimes in the} presence of ordinary matter distribution
at the throat. One may also find further investigations on BD wormholes in the literature, see e.g.,~\cite{bd3,bd3rev,bd5,bd6,bd7}. Motivated by these considerations, in the present work we wish to build and study wormhole solutions in BD theory in the presence of matter distribution with an anisotropic EMT. The article is then organized as follows: In Sec.~(\ref{BDFES}) we provide a brief review on the field equations of BD theory. In Sec.~(\ref{wormsolssec}), using the general static spherically symmetric MT metric, we find the corresponding differential equations to be solved subject to the physical conditions on wormhole geometry. In Sec.~(\ref{redzero}) we present exact zero tidal force wormhole solutions and examine the energy conditions for them. Sect.~(\ref{rednzero}) is devoted to the solutions with non-vanishing redshift function. Some attributes of the obtained solutions are discussed in Sec.~(\ref{sofuis}) and Sec.~(\ref{concluding}) presents concluding remarks. We use a system of units so that $\hbar=c=G=1$.
\section{Field Equations of Brans-Dicke Theory}\label{BDFES}
Motivated by previous works of Pascual Jordan in 1950s~\cite{Jordan1950}, in 1961, Robert H. Dicke and Carl Brans introduced a gravitational theory later known as the BD theory~\cite{BDTH}. The essential feature of the BD theory is the presence of a scalar field that together with the spacetime metric describe gravitational interaction. This alternative theory of gravity combines the equivalence principle, a cornerstone of GR, with Mach's principle, which states that \lq{}\lq{}the universal gravitational constant should be a function of the mass distribution of the Universe\rq{}\rq{}~\cite{maeda2003}. In a four-dimensional Riemannian spacetime endowed with the metric $g_{\mu\nu}$, a metric compatible connection ($\nabla_\alpha g_{\mu\nu}=0$), and the scalar field $\phi$, the BD action in Jordan frame is expressed as~\cite{Faraoni}
\bea {\mathcal S}=\f{1}{16\pi}\int  d^4x\sqrt{-g}\left[\phi R-\f{\omega}{\phi}g^{\mu\nu}\nabla_\mu\phi\nabla_\nu\phi-V(\phi)\right]+{\mathcal S}^{({\rm m})},\label{actionbd}
\eea
where
\bea 
{\mathcal S}^{(m)}=\int  d^4x\sqrt{-g}\mathcal{L}^{(m)},
\eea
is the matter sector of the action, $R$ is the Ricci scalar and $\omega$ is a dimensionless real constant called BD coupling parameter. The matter part is not directly coupled to the BD scalar field since $\mathcal{L}^m=\mathcal{L}^m(\Psi,\partial\Psi,x^\mu)$, where $\Psi$ represents matter fields, i.e., fields other than the BD scalar field $\phi$. However, in the gravitational sector, the scalar field is non-minimally coupled to {Ricci scalar}. This makes the BD scalar field to act as a source of gravitational field also, since it is not of type of material fields. {The scalar field potential $V(\phi)$ represents the energy associated with the scalar field and the dynamics of the scalar field is governed by an evolution equation that incorporates the potential. In cosmological scenarios the specific form of $V(\phi)$ dictates how the scalar field evolves over time, influencing cosmological dynamics and structure formation~\cite{Faraoni},\cite{Singh1983}.} Taking variation of the action Eq.~(\ref{actionbd}) with respect to metric we obtain the BD field equation as
\bea\label{bdfield}
{\rm G}_{\mu\nu}=\f{8\pi}{\phi}{\rm T}^{\rm (m)}_{\mu\nu}+\f{\omega}{\phi^2}\Big[\nabla_\mu\phi\nabla_\nu\phi-\f{1}{2}g_{\mu\nu}\nabla^\alpha\phi\nabla_\alpha\phi\Big]+\f{1}{\phi}\left(\nabla_\mu\nabla_\nu\phi-g_{\mu\nu}\Box\phi\right)-\f{V(\phi)}{2\phi}g_{\mu\nu},
\eea
where
\bea\label{emt}
{\rm T}^{\rm (m)}_{\mu\nu}=-\f{2}{\sqrt{-g}}\f{\delta}{\delta g^{\mu\nu}}\left(\sqrt{-g}\mathcal{L}^{\rm (m)}\right),
\eea
is the matter EMT. Varying the action Eq.~(\ref{actionbd}) with respect to $\phi$ gives the evolution equation for BD scalar field as
\be\label{bdevolve}
\f{2\omega}{\phi}\Box\phi+R-\f{\omega}{\phi^2}\nabla^\alpha\phi\nabla_\alpha\phi-\f{dV}{d\phi}=0.
\ee
From Eq.~(\ref{bdfield}) we get the following expression for Ricci scalar
\be\label{bdevolve1}
R=-\f{8\pi}{\phi}{\rm T}^{(m)}+\f{\omega}{\phi^2}\nabla^\alpha\phi\nabla_\alpha\phi+\f{3\Box\phi}{\phi}+\f{2V}{\phi},
\ee
where ${\rm T}^{\rm (m)}$ is the trace of EMT. Using the above equation to eliminate $R$ from Eq.~(\ref{bdevolve}) we finally get the evolution equation for BD scalar field as
\be\label{bdevol}
(2\omega+3)\Box\phi-8\pi {\rm T}^{(m)}-\phi\f{dV}{d\phi}+2V=0.
\ee
The above equation along with Eq.~(\ref{bdfield}) construct the basic field equations of the BD theory. These tensor equations will be reduced to a system of coupled second-order non-linear differential equations upon considering the well-known MT line element~\cite{mt}. This is the subject of the next section where we shall deal with this system by adopting a suitable strategy for solving the differential equations. {It is noteworthy that the BD coupling parameter determines the strength of nonminimal coupling between the scalar field and the spacetime curvature. The larger the value of the coupling parameter, the smaller the effects of the scalar field, and in the limit $\omega\rightarrow\infty$, the theory becomes indistinguishable from GR in all its predictions~\cite{maeda2003},\cite{CMWill}. This parameter is constrained by observational tests, particularly in weak-field regimes. For example, Solar system experiments, such as the Cassini mission's measurement of the Shapiro time delay, predict the lower bound $\omega>40000$~\cite{CMWill},\cite{Berto2003}. However, this parameter may be less than this value typically as large as 500 and 1000 on cosmological scales~\cite{omega500} or even it may assume negative values such as $\omega<-120$ obtained from WMAP and SDSS data~\cite{omegam120}. Gravitational wave observations of inspiraling compact binaries such as a neutron star and a black hole could give the constraint $\omega\gtrsim10$~\cite{nsbhgw}, see also~\cite{nsbhgw1} for further updates. While large values of BD coupling parameter recovers GR, small values enhances deviations, potentially violating energy conditions thus enabling exotic spacetimes~\cite{bd,bd1,bd2,bd4}. In cosmological contexts, BD theories with $\omega\approx-3/2$ provide a border between a standard scalar field model and a ghost/phantom model~\cite{Dabrow2007}, highlighting the parameter's role in linking modified gravity theories to dark energy models and quantum corrections~\cite{Faraoni}.} In this study, we exclude the case for which $\omega=-3/2$ since for this value of coupling parameter, the BD theory exhibits a pathological behavior in the way that the scalar field is non-dynamical and the Cauchy problem is ill-posed~\cite{BDpath},\cite{omega32}. 
\section{Spherically Symmetric Wormholes}\label{wormsolssec}
In the present section we search for a class of wormhole configurations in BD theory whose supporting matter obeys a linear EoS between radial pressure and energy density. The general static and spherically symmetric line element representing a wormhole spacetime is given by~\cite{mt}
\begin{align}\label{metric}
	ds^2=-{\rm e}^{2\Phi(r)}dt^2+\left(1-\f{b(r)}{r}\right)^{-1}dr^2+r^2d\Omega^2,
\end{align} 
where $d\Omega^2=d\theta^2+\sin^2\theta d\varphi^2$ is the standard line element on a unit two-sphere {and,} $\Phi (r)$ and $b(r)$ are denoted as redshift and shape functions, respectively. {The redshift function, determines the gravitational redshift experienced by light or the magnitude of tidal forces experienced by a timelike traveler passing through the wormhole and influences the rate at which the time flows at different radial positions~\cite{mt},\cite{GrMTW}. The shape function determines the spatial geometry of the wormhole, particularly the radius of the throat, how it flares outward and how the wormhole's cross-section varies with radial distance~\cite{khu,mt}. In BD theory, the scalar field acts as an additional degree of freedom (source of exotic matter) that affects the behavior of the redshift and shape functions through the BD field equations. This helps to construct wormhole solutions where normal matter threading the wormhole geometry respects the NEC throughout the geometry~\cite{LOBOBDEXOT2011}.} The radial coordinate lies in the interval $r\in[r_0,\infty)$ where the surface at $r=r_0$ is called the wormhole\rq{}s throat. {Let us now} define the time-like and space-like vector fields, respectively as $u^\mu=[{\rm e}^{-\Phi(r)},0,0,0]$ and $v^\alpha=\left[0,\sqrt{1-b(r)/r},0,0\right]$, so that $u^\mu u_\mu=-1$ and $v^\alpha v_\alpha=1$. The anisotropic {\rm EMT} of matter source then takes the following form
\be\label{emtaniso}
{\rm T}^{(m)}_{\mu\nu}=[\rho(r)+p_t(r)]u_\mu u_\nu+p_t(r)g_{\mu\nu}+[p_r(r)-p_t(r)]v_\mu v_\nu,
\ee
where $\rho(r)$ is the energy density and $p_r(r)$ and $p_t(r)$ are {the pressure profiles in radial and tangential directions}, respectively. The components of field equation (\ref{bdfield}) then read
\bea
16\pi\rho(r)&=&\f{2\phi b^\prime}{r^2}-2\left(1-\f{b}{r}\right)\phi^{\prime\prime}-\f{\omega}{\phi}\left(1-\f{b}{r}\right)(\phi^\prime)^2+\f{\phi^\prime}{r}\left(b^\prime+3\f{b}{r}-4\right)-V(\phi),\label{rhoex}\\
16\pi p_r(r)&=&-\f{\omega}{\phi}\left(1-\f{b}{r}\right)(\phi^\prime)^2+2\left(1-\f{b}{r}\right)\left(\Phi^\prime+\f{2}{r}\right)\phi^\prime+4\phi\left(1-\f{b}{r}\right)\f{\Phi^\prime}{r}-2\f{\phi b}{r^3}+V(\phi),\label{prexp}\\
16\pi p_t(r)&=&2\left(1-\f{b}{r}\right)\Phi^{\prime\prime}+2\phi\left(1-\f{b}{r}\right)\Phi^{\prime\prime}+\f{\omega}{\phi}\left(1-\f{b}{r}\right)(\phi^\prime)^2+\f{2}{r}\left[(r-b)\Phi^\prime-\f{b^\prime}{2}+1-\f{b}{2r}\right]\phi^\prime\nn&+&2\phi\left(1-\f{b}{r}\right)(\Phi^\prime)^2-\f{\phi}{r}\left(b^\prime-2+\f{b}{r}\right)\Phi^\prime-\f{b^\prime}{2r^2}+\f{b}{2r^3}+\f{V(\phi)}{2\phi},\label{ptexp}
\eea
where $\prime\equiv d/dr$. Also the evolution equation for BD scalar field, Eq.~(\ref{bdevol}), takes the form
\be\label{bdevol1}
\left(1-\f{b}{r}\right)\phi^{\prime\prime}+\left[(r-b)\Phi^\prime-\f{b^\prime}{2}-\f{3b}{2r}+2\right]\f{\phi^\prime}{r}+\f{1}{2\omega+3}\left[8\pi(\rho-p_r-2p_t)+2V-\phi\f{dV}{d\phi}\right]=0.
\ee
The {equation that governs the {\rm EMT} conservation}, i.e., $\nabla^\mu {\rm T}^{\rm (m)}_{\mu\nu}=0$ {leaves us with the following expression}
\be\label{conseq}
\Phi^\prime(\rho+p_r)+p_r^\prime+\f{2}{r}\left(p_r-p_t\right)=0.
\ee
Considering the above system of differential equations we can figure out that substituting for energy density and pressure profiles from Eqs.~(\ref{rhoex})-(\ref{ptexp}) into the conservation equation (\ref{conseq}) and simplifying the result, we arrive at evolution equation for BD scalar field, i.e., Eq.~(\ref{bdevol1}). Moreover, substituting for the energy density and radial pressure into Eqs.~(\ref{bdevol1}) and (\ref{conseq}) along with omitting $V^\prime(r)$, one readily arrives at the expression given for tangential pressure, i.e., Eq.~(\ref{ptexp}). This means that, from equations (\ref{rhoex})-(\ref{conseq}) only four of them are independent. We therefore take equations~(\ref{rhoex}) and (\ref{prexp}) along with (\ref{bdevol1}) and (\ref{conseq}) as independent ones to be solved for possible wormhole solutions. For the linear EoS $p_r(r)=w\rho(r)$, equation~(\ref{conseq}) gives the tangential pressure as\footnote{{The adoption of a linear EoS in wormhole solutions is motivated by several theoretical and practical considerations. Nevertheless, it is possible to consider more complicated forms for the equation of state e.g., polytropic $p=K\rho^{1+\f{1}{n}}$~\cite{polyeos,polyeos1}, quadratic $p=\alpha\rho+\beta\rho^2$~\cite{quadeos}, Chaplygin $p=-A/\rho$ and generalized Chaplygin gas $p=-A/\rho^\alpha$~\cite{polchapeos}, Van der Waals $p=\gamma\rho/(1-\beta\rho)-\alpha\rho^2$~\cite{vandereos} and Logarithmic $p=A\ln(\rho/\rho_0)$~\cite{logeos}. The first reason for choosing a linear EoS is that, it provides a mathematically tractable framework that allows for exact analytical solutions to the field equations of a gravitational theory, while nonlinear EoS would often lead to complicated set of equations that require numerical considerations. The second reason is that, a linear EoS supports basic phenomenological behaviors, e.g., the EoS parameters for dust and radiation fluids are $w=0$ and $w=1/3$ respectively. In $\Lambda${\rm CDM} model, a cosmological constant is described by dark energy with EoS parameter $w\approx-1$~\cite{coscons}. Also in canonical scalar field models, quintessence and phantom fields are described by EoS parameters $-2/3\leq w<-1/3$~\cite{quint} and $w<-1$~\cite{darkphantom}, respectively. The third reason is that, a linear EoS aligns with a plenty of well-known approaches in wormhole physics, where the simplicity of dealing with the field equations of GR or other modified gravities provides a tractable way to analyze the exotic matter required to sustain a wormhole throat. Moreover a linear EoS is a useful assumption which facilitates comparisons with known cosmological and astrophysical fluids (e.g., dark and phantom energies and quintessence) that may mimic wormhole supporting matter, see~\cite{FLoboBook,hisworm,lobocgqreview},\cite{phantworm,intdarksec},\cite{phantwormhool} and discussions therein.}}
\be\label{tanpres}
p_t(r)=\f{r}{2}\Phi^\prime(r)\rho(r)(1+w)+\f{w}{2}\left[2\rho(r)+r\rho^\prime(r)\right],
\ee
whereby substituting for the tangential pressure into Eq.~(\ref{bdevol1}) we get
\bea\label{bdevoltov}
&&\phi^\prime\left[\left(1-\f{b}{r}\right)\Phi^\prime-\f{b^\prime}{2r}+\f{2}{r}-\f{3b}{2r^2}\right]-\f{8}{2\omega+3}\left[\pi\rho(w+1)r\Phi^\prime+w\pi r\rho^\prime+\pi\rho(3w-1)-\f{V}{4}\right]\nn
&&-\f{\phi V^\prime}{(2\omega+3)\phi^\prime}+\phi^{\prime\prime}\left[1-\f{b}{r}\right]=0.
\eea
Also from Eqs.~(\ref{rhoex}) and (\ref{prexp}) we have
\bea\label{bdrhowpr}
&&\f{w}{8\pi}\left[1-\f{b}{r}\right]\phi^{\prime\prime}+\f{\phi^\prime}{8\pi}\left(\Phi^\prime\left[1-\f{b}{r}\right]-\f{wb^\prime}{2r}-\f{b}{2r^2}(3w+4)+\f{2}{r}(w+1)\right)+\nn&&\f{\omega(w-1)}{16\pi}\left[1-\f{b}{r}\right]\f{\phi^{\prime2}}{\phi}
+\f{\phi}{4\pi r}\left[1-\f{b}{r}\right]\Phi^\prime-\f{w\phi b^\prime}{8\pi r^2}-\f{b\phi}{8\pi r^3}+\f{V}{16\pi}(w+1)=0.
\eea
Equations (\ref{bdevoltov}) and (\ref{bdrhowpr}) construct a system of two coupled differential equations to be solved for five unknowns $\{b(r),\phi(r),\Phi(r),V(r),\rho(r)\}$. This system is under-determined as the number of independent equations are less than the unknowns, therefore, in order to solve it we have to specify the functionality of three of the unknowns. Let us proceed with considering the term given in the square brackets of Eq.~(\ref{bdevoltov}) from which we can find the following differential equation for energy density
\be\label{diffeqrhof}
\pi\rho(w+1)r\Phi^\prime+w\pi r\rho^\prime+\pi\rho(3w-1)-\f{V}{4}-f(r)=0,
\ee
where we have introduced $f(r)$ as a well-behavior auxiliary function of radial coordinate that will be useful to find consistent wormhole solutions. The solution then reads
\bea\label{rhosol}
\rho(r)=r^{\f{1}{w}-3}\left\{\f{1}{4\pi w}\int\left[V(r)+4f(r)\right]r^{\f{2w-1}{w}}{\rm exp}\left[\f{1+w}{w}\Phi(r)\right]dr+{\rm C}_1\right\}{\rm exp}\left[-\f{1+w}{w}\Phi(r)\right],
\eea
where ${\rm C}_1$ is an integration constant to be determined later. Substituting the above solution back into Eq.~(\ref{bdevoltov}) we get
\bea\label{bdtovrhosol}
\phi^{\prime\prime}\left[1-\f{b}{r}\right]+\phi^\prime\left[\left(1-\f{b}{r}\right)\Phi^\prime-\f{b^\prime}{2r}+\f{2}{r}-\f{3b}{2r^2}\right]-\f{\phi V^\prime}{(2\omega+3)\phi^\prime}-\f{8f}{2\omega+3}=0.
\eea
From now on we have a system of two coupled differential equations, i.e., Eqs.~(\ref{bdrhowpr}) and (\ref{bdtovrhosol}) to be solved for the five unknowns $\{b(r),\phi(r),\Phi(r),V(r),f(r)\}$. In the next section we try to build and study physically reasonable wormhole solutions for which the density profile obeys Eq.~(\ref{rhosol}) along with assuming specific forms for the unknown function $f(r)$, the BD scalar field $\phi(r)$ and the redshift function $\Phi(r)$. {It should be noted that when exploring wormhole solutions in GR or alternative theories, the choice between constant and variable redshift functions depends on specific modeling objectives. These objectives can be defined as traversability which ensures the wormhole is traversable by humans or particles (no horizons, tolerable tidal forces), satisfaction or at least minimal violation of NEC and WEC, assymptotic flatness, observational signatures such as lensing effects and stability of the wormhole configuration.}
\section{Solutions with $\Phi(r)=0$}\label{redzero}
\begin{figure}
	\begin{center} 
		\includegraphics[width=7.7cm]{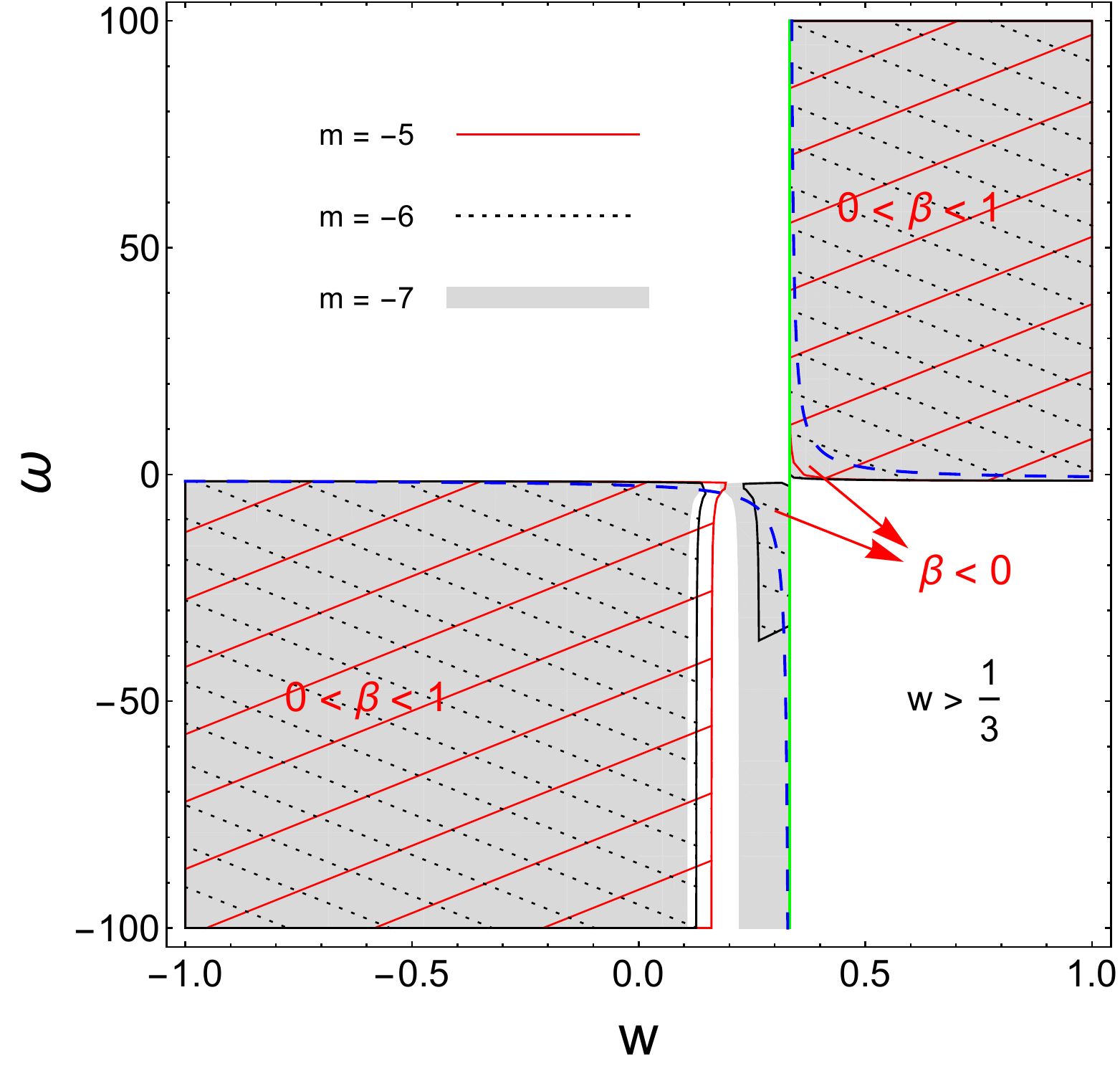}
		\includegraphics[width=7.7cm]{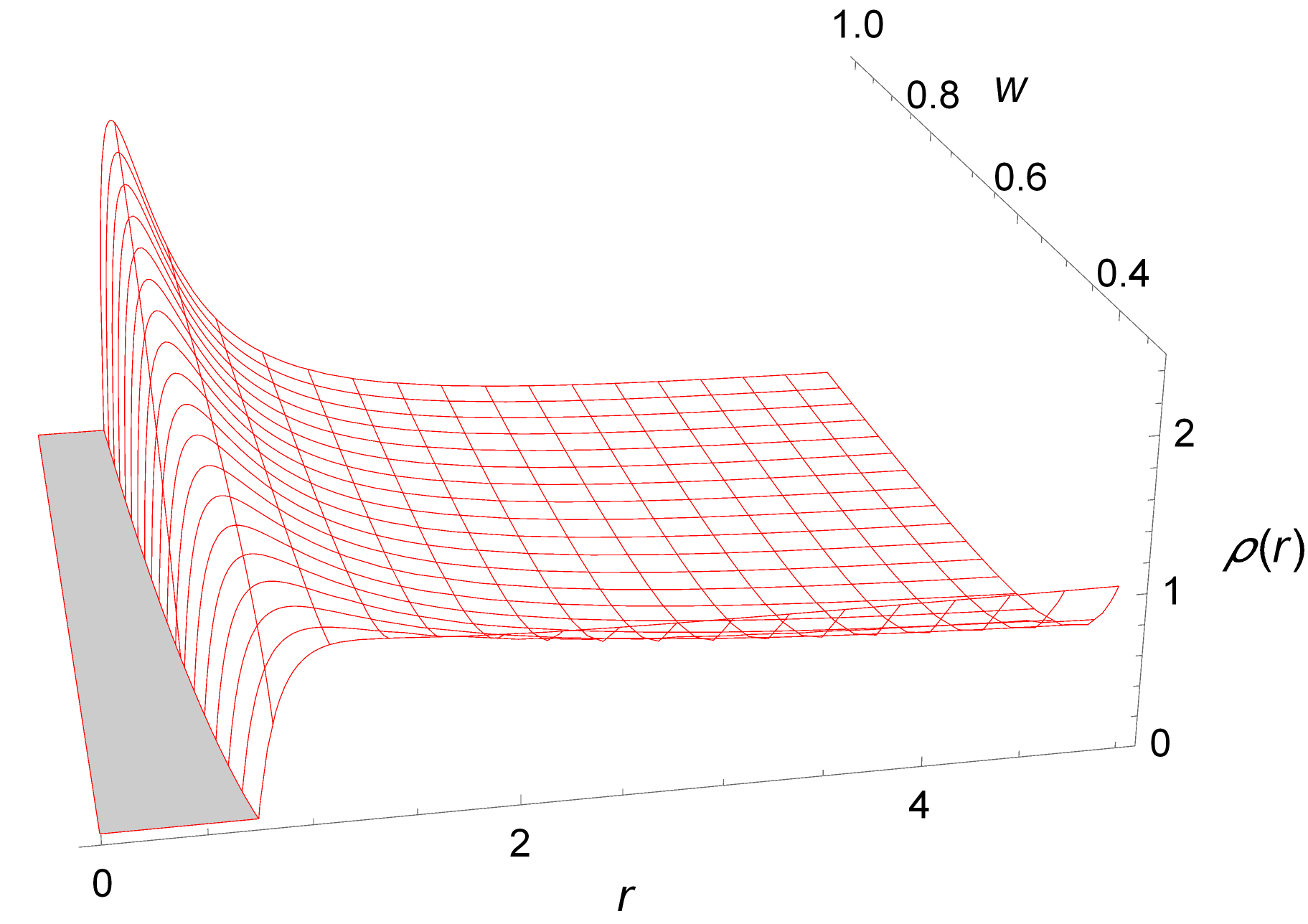}
		\caption{Left panel: the allowed values of EoS and BD coupling parameters for $\rho_0=1$, $r_0=1$ and different values of the exponent $m$. The green vertical line separates the region with $w>1/3$ which corresponds to asymptotically flat solutions. Right panel: Behavior of energy density against radial coordinate for different values of EoS parameter. For the model parameters we have set $\rho_0=r_0=1$, $m=-5$ and $\omega=60$.}\label{fig1}
	\end{center}
\end{figure}

\begin{figure}
	\begin{center} 
		\includegraphics[width=7.7cm]{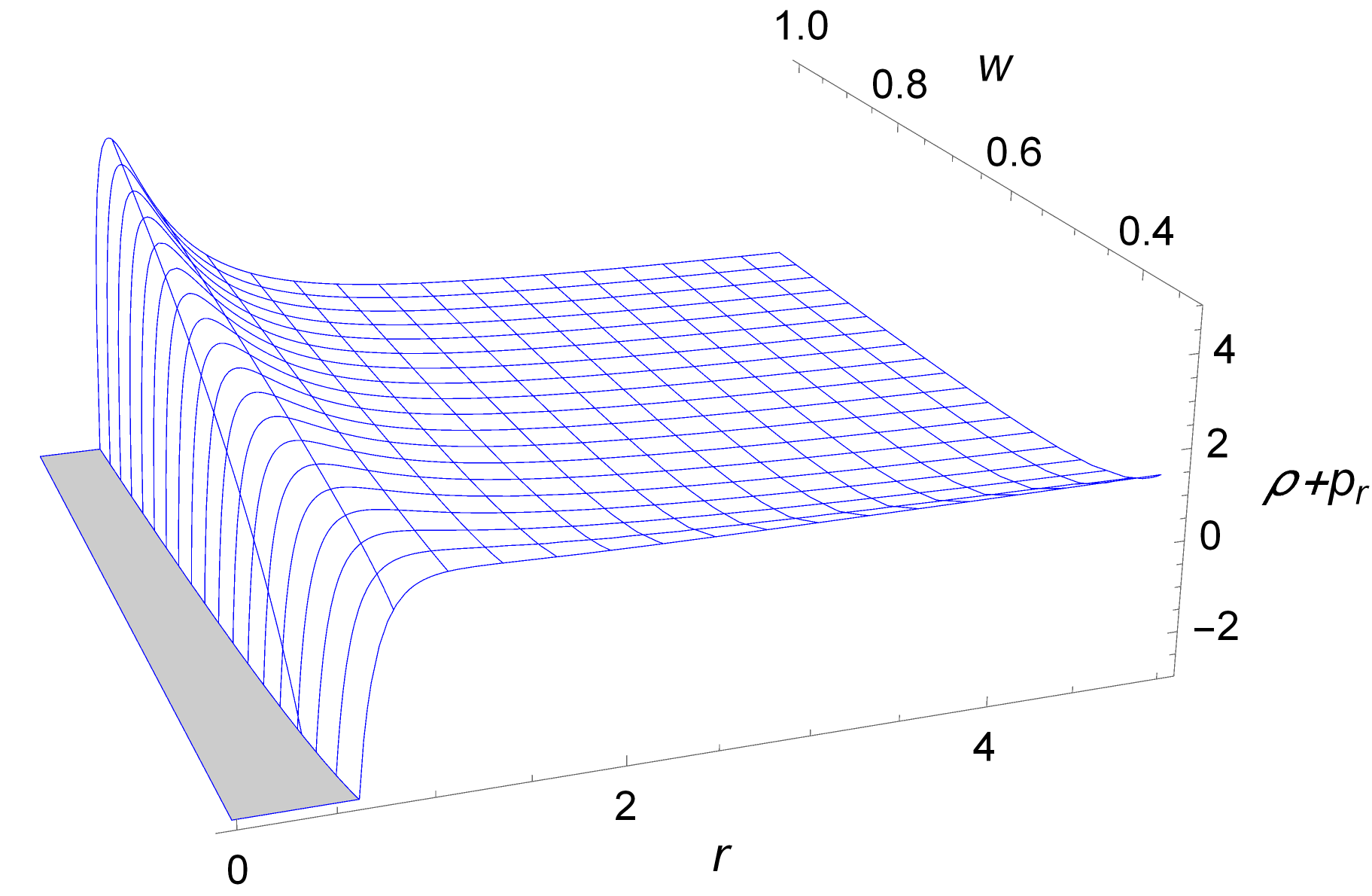}
		\includegraphics[width=7.7cm]{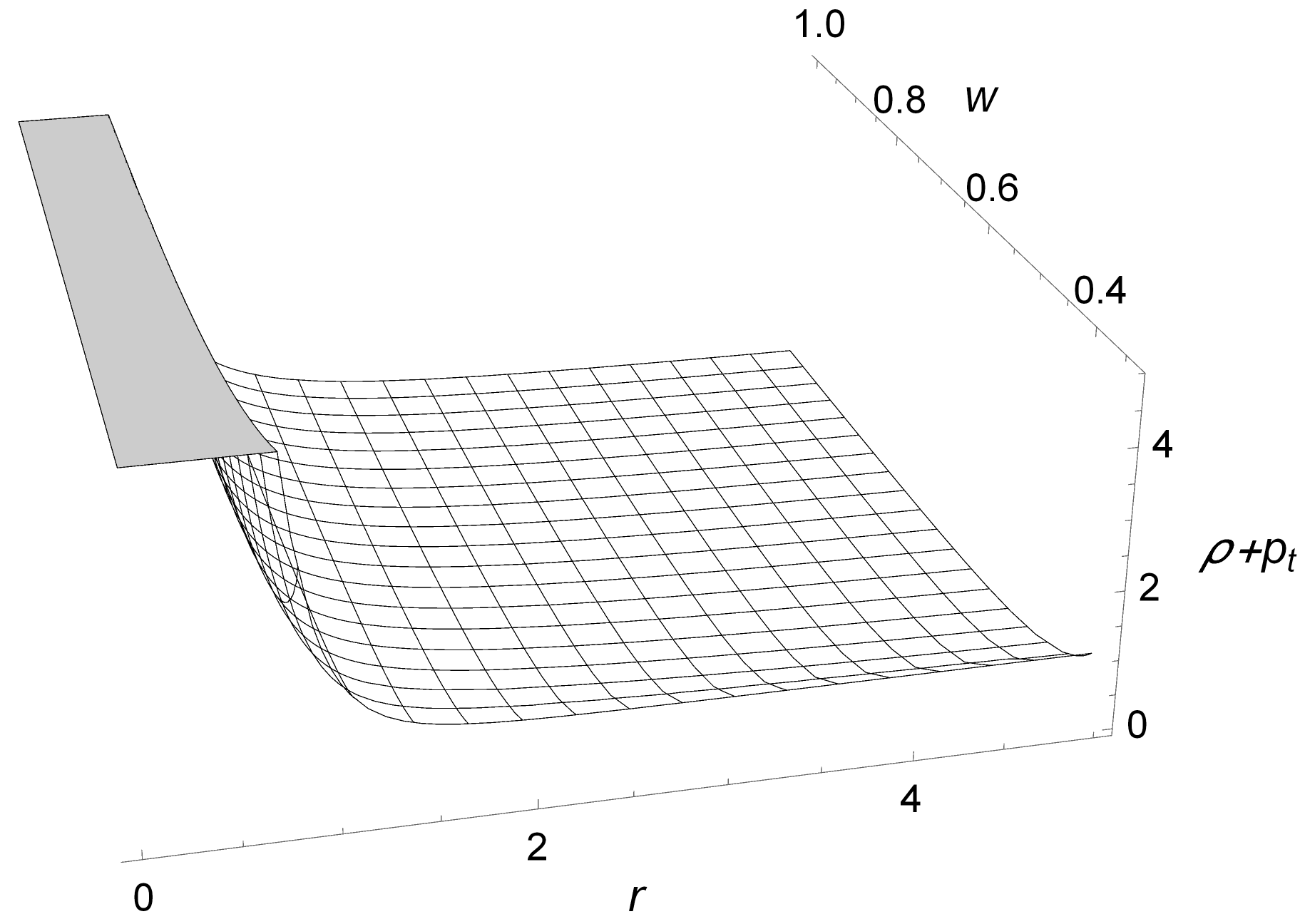}
		\caption{Radial (left panel) and tangential (right panel) components of NEC against radial coordinate for different values of EoS parameter. For the model parameters we have set $\rho_0=r_0=1$, $m=-5$ and $\omega=60$.}\label{fig2}
	\end{center}
\end{figure}

\begin{figure}
	\begin{center} 
		\includegraphics[width=7.7cm]{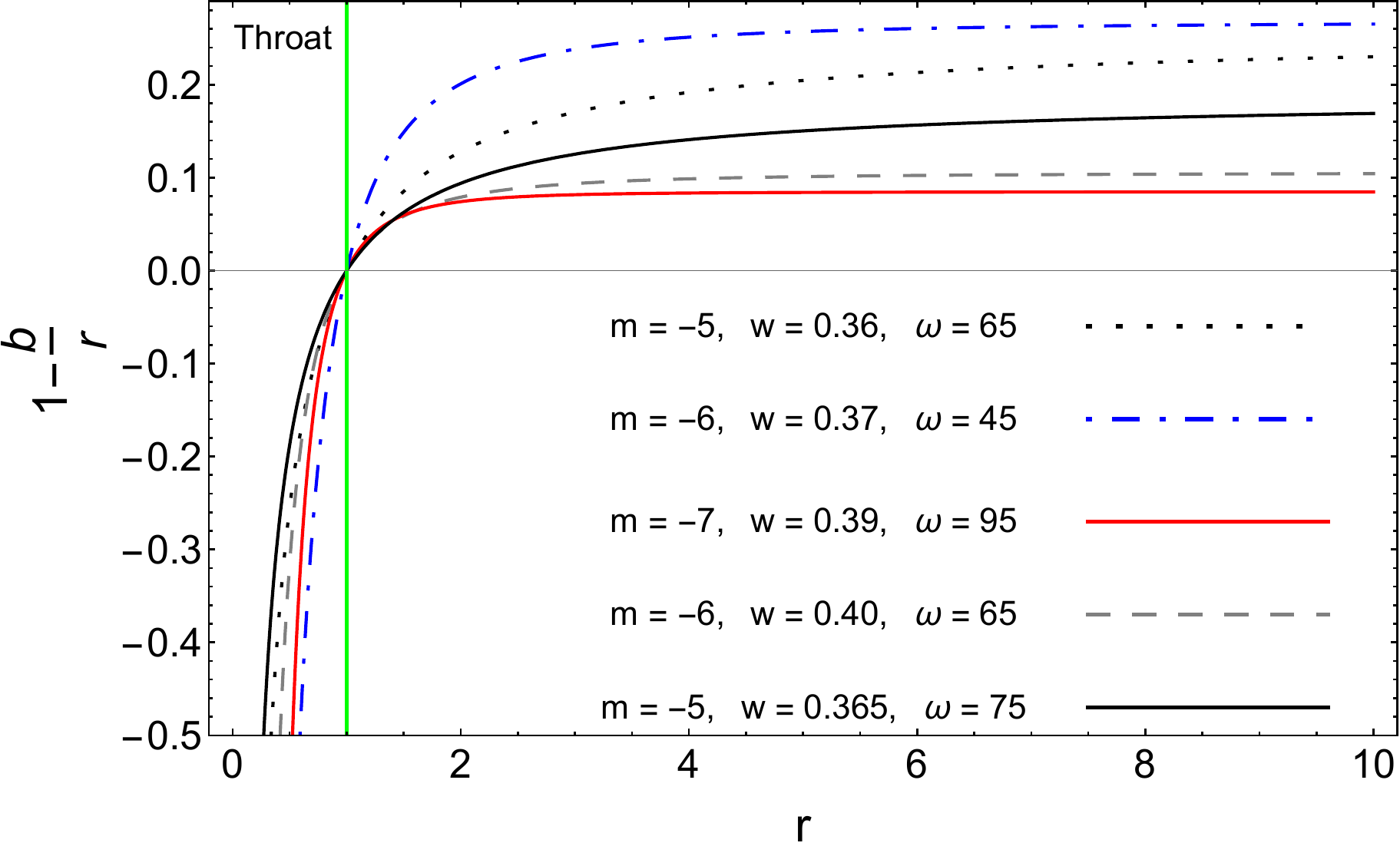}
		\includegraphics[width=7.7cm]{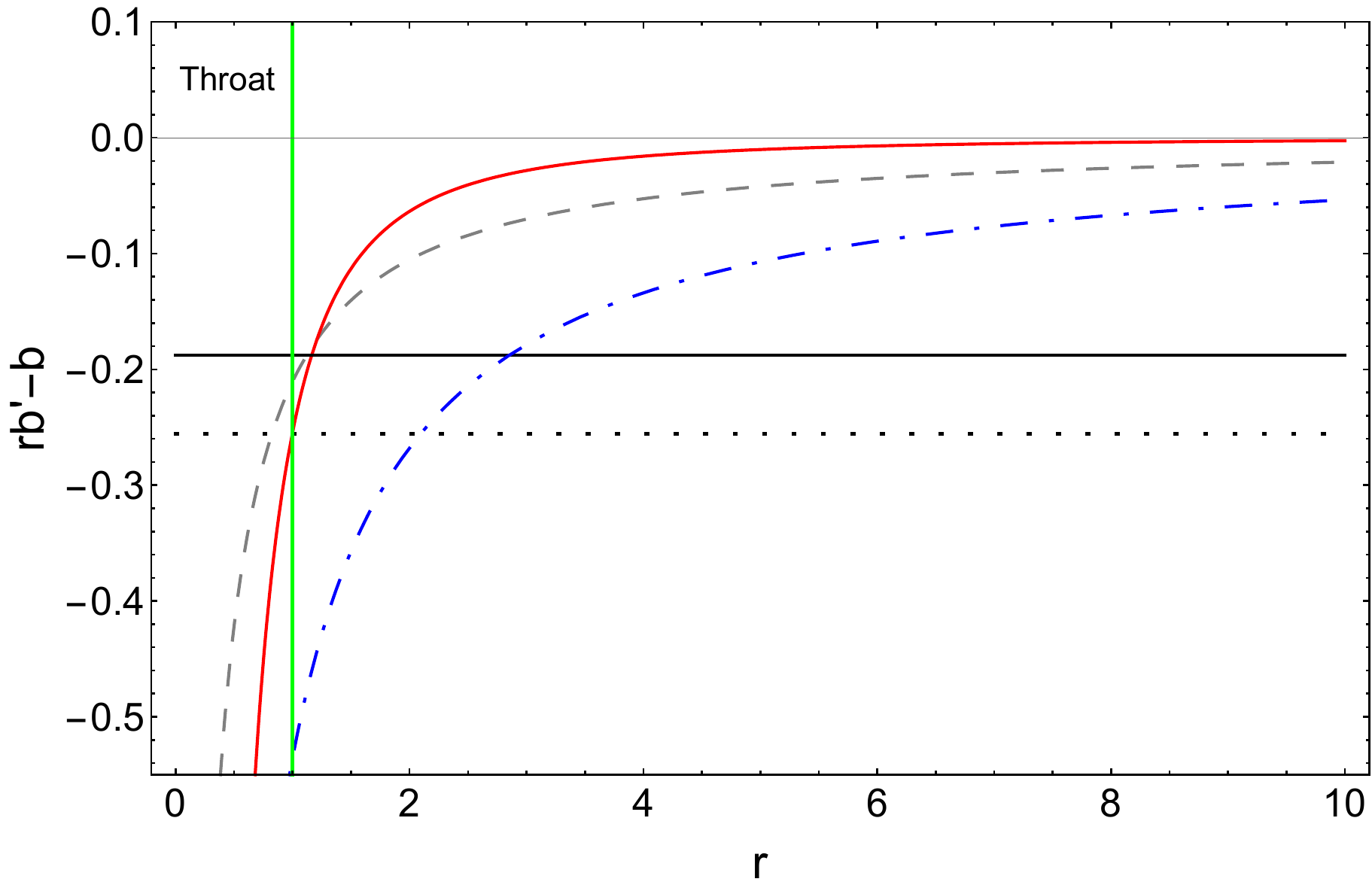}
		\caption{Left panel: The behavior of $g_{rr}^{-1}$ for the allowed values of model parameters subject to the left panel of Fig.~(\ref{fig1}). Right panel: Flare-out condition against radial coordinate for the same values of model parameters as of the left panel. We have set the throat radius as $r_0=1$.}\label{fig3}
	\end{center}
\end{figure}

In this section we present zero tidal force wormhole solutions whose supporting matter obeys a linear EoS between radial pressure and energy density. {It should be noted that for a traversable wormhole no horizon should exist throughout the spacetime}, since the presence of horizon, defined as the surface with ${\rm e }^{2\Phi(r)}\rightarrow0$, would prevent round trips through the wormhole; {Therefore}, the redshift function must be finite {throughout the wormhole spacetime. In order that the shape function describes} a physically viable wormhole solution it has to fulfill the following fundamental conditions~\cite{mt,mt1}
\begin{enumerate}[label=(\roman*)]
	\item $\left(1-\f{b(r)}{r}\right)\Big|_{r=r_0}=0$: This gives the radius of the throat at $r=r_0$.\label{c1}
	\item $rb^\prime(r)-b(r)<0$ for $r>r_0$: This is the fundamental flare-out condition which at the throat reads, $b^\prime(r_0)<1$.\label{c2}
	\item $\lim\limits_{r\rightarrow\infty}\f{b(r)}{r}=0$: This condition gives the asymptotic flatness of the metric.\label{c3}
	\item $\f{b(r)}{r}<1$ or equivalently $g_{rr}^{-1}(r)>0$ for $r>r_0$: This condition guarantees that the (Lorentzian) metric signature is preserved.\label{c4}
\end{enumerate}
{Next, we proceed to obtain} the shape function and scalar field potential, assuming the functionality of BD scalar field as $\phi(r)=\phi_0\left(\f{r_0}{r}\right)^n$ where $n\in\mathbb{R}^+$ and $\phi_0$ is the value of scalar field at the throat. Moreover we set $f(r)=r^m$ where $m\in\mathbb{R}$. {For the system (\ref{bdrhowpr}) and (\ref{bdtovrhosol}) we then find} the following exact {solutions} for $n=2$, as
\bea
b(r)\!\!\!\!&=&\!\!\!\!b_1r+b_2r^{m+5}+b_3\left(\f{r_0}{r}\right)^{\f{1-4w}{w}},\label{solb}\\
V(r)\!\!\!\!&=&\!\!\!\!V_1r^m+\f{V_2}{r^4}+V_3\left(\f{r_0}{r}\right)^{\f{1+w}{w}},\label{solV}
\eea
where
\bea\label{bha}
b_1&=&\f{[w(6\omega+7)-2\omega-5][1+w(m+1)]}{(3w-1)(2\omega+3)[1+w(m+1)]},~~~~b_2=\f{\left(12w^2+8w-4\right)}{(3w-1)[1+w(m+1)](2\omega+3)\phi_0r_0^2},\nn b_3&=&\f{2(w+1)\left[2(1-3w)r_0^{m+3}+r_0(1+w(m+1))\phi_0\right]}{(3w-1)[1+w(m+1)](2\omega+3)\phi_0},~~~~V_1=\f{8(3w^2-4w+1)}{(3w-1)[1+w(m+1)]},\nn V_2&=&\f{2(w+1)[1+w(m+1)]\phi_0r_0^2}{(3w-1)[1+w(m+1)]},~~~~V_3=\f{4(w-1)\left[2(1-3w)r_0^{m+2}+(1+w(m+1))\phi_0\right]}{(3w-1)[1+w(m+1)]r_0^2},\nn
\eea
and the integration constant has been set according to the condition~\ref{c1}. From Eq.~(\ref{solb}) the flare-out condition~\ref{c2} at the throat leads to the following inequality
\bea\label{flr0}
rb^\prime-b<0\Big|_{r=r_0}\Rightarrow b_1+(m+5)b_2r_0^{m+4}+\f{(4w-1)}{wr_0}b_3<1.
\eea
Moreover, condition~\ref{c3} on asymptotic flatness, requires that $w<1/3$ and $m<-4$. Regarding condition~\ref{c4}, in order that $b(r)/r<1$ for $r>r_0$ one may require that this ratio be a decreasing function of radial coordinate. Hence, the slope of the function $b(r)/r$ should be negative for $r\geq r_0$, this gives
\be\label{mosht}
\f{d}{dr}\left(\f{b(r)}{r}\right)\Bigg|_{r=r_0}\!\!\!\!\!\!<0~~~~~~\Rightarrow~~~~~~\f{1}{r^2}\left[rb^\prime(r)-b(r)\right]\Big|_{r=r_0}\!\!\!\!<0,
\ee
which is nothing but the condition \ref{c2}. As stated before, in the context of GR traversability of a wormhole configuration requires the violation of NEC, which in turn, signals the existence of exotic matter with negative energy density. In other words, the feature of wormhole traversability is accompanied by restrictions on the EMT components that must be respected at or near any wormhole throat. Consider a congruence of null rays defined by the null vector field $k^\mu$ with $k^\mu k_\mu=0$~\cite{PoissonBook}, the traversability of wormhole then leads to the condition ${\rm T}_{\mu\nu}k^\mu k^\nu\leq0$ which implies the violation of NEC by the EMT of an exotic matter~\cite{lobocgqreview},\cite{Hochberg1998}. For the {\rm EMT} defined in Eq.~(\ref{emtaniso}) the {\rm NEC} is expressed as
\bea
&&\rho(r)+p_{r}(r)\geq0,~~~~~~~~\rho(r)+p_{t}(r)\geq0\label{nec}.
\eea
{Also, in order to construct physically reasonable wormhole geometries, one may require that the obtained solutions} respect the WEC. According to this condition, {the energy density as measured by} any observer {in the spacetime has to be} non-negative or {equivalently,} ${\rm T}_{\mu\nu}u^\mu u^\nu\geq0$ {with $u^\mu$ being a timelike 4-vector field}~\cite{PoissonBook}. The {\rm WEC} then leads to the following inequalities
\bea
&&\rho(r)\geq0,~~~~~~~~\rho(r)+p_{r}(r)\geq0,~~~~~~~~\rho(r)+p_{t}(r)\geq0.\label{wec}
\eea
We {also} note that {\rm WEC} implies the null form. For the solutions obtained in Eqs.~(\ref{solb}) and (\ref{solV}) we can find the energy density, Eq.~(\ref{rhosol}), and pressure profiles as
\bea
&&\rho(r)=\rho_0\left(\f{r_0}{r}\right)^{\f{3w-1}{w}}+\rho_1\left(\f{r_0}{r}\right)^{\f{1+w}{w}}+\rho_2r^m+\f{\rho_3}{r^4},~~~~~~~~~pr(r)=w\rho(r),\label{energy}
\\
&&p_{t}(r)=p_1\left(\f{r_0}{r}\right)^{\f{1+w}{w}}+p_2r^m+\f{p_3}{r^4},\label{presst}
\eea
where the integration constant has been set so that $\rho(r_0)=\rho_0$ and
\bea\label{densities}
\rho_1\!\!\!\!&=&\!\!\!\!\f{\phi_0[1+w(m+1)]-2(3w-1)r_0^{m+2}}{2\pi(3w-1)[1+w(m+1)]r_0^2},~~~~\rho_2=\f{1}{\pi[1+w(m+1)]},~~~~\rho_3=\f{\phi_0r_0^2}{2\pi(1-3w)},\nn p_1\!\!\!\!&=&\!\!\!\!\f{[w^2-4(2\omega+3)w-1]\left[2(1-3w)r_0^{m+2}+\phi_0(1+w(m+1))\right]}{8\pi r_0^2w(2\omega+3)(3w-1)[1+w(m+1)]},\nn p_2\!\!\!\!&=&\!\!\!\!\f{2[w(m+2)+m-8\omega-10]}{4\pi(2\omega+3)[1+w(m+1)]},~~~~~~~p_3=\f{(6\omega+11)(w+1)\phi_0r_0^2}{8\pi(2\omega+3)(3w-1)}.
\eea
From the above expressions we find
\bea
\rho(r)+p_{r}(r)=(1+w)\rho(r),~~~~~~~\rho(r)+p_{t}(r)&=&\rho_0\left(\f{r_0}{r}\right)^{\f{3w-1}{w}}+(\rho_1+p_1)\left(\f{r_0}{r}\right)^{\f{1+w}{w}}+(\rho_2+p_2)r^m\nn&+&\f{\rho_3+p_3}{r^4},\label{energypt}
\eea
Hence, the energy conditions at the throat take the form
\bea
\rho\Big|_{r=r_0}\!\!\!\!\!&=&\rho_0\geq0,~~~~~~~~~\rho+p_{r}\Big|_{r=r_0}\!\!=(1+w)\rho_0\geq0,\nn\rho+p_{t}\Big|_{r=r_0}\!\!\!\!&=&\rho_0+\rho_1+p_1+(\rho_2+p_2)r_0^m+\f{\rho_3+p_3}{r_0^4}\geq0.\label{enprpt0}
\eea
From the first part of Eq.~(\ref{enprpt0}) we observe that the positivity of energy density at wormhole throat requires that $\rho_0>0$. From the second part we find that at the throat, the EoS must obey $w\geq-1$ so that the WEC in radial direction is preserved. From Eq.~(\ref{solb}) and first part of Eq.~(\ref{energy}) we deduce that the condition $w<1/3$ leads to $\rho(r)\rightarrow\infty$ as $r\rightarrow\infty$ which is clearly inconsistent with asymptotic flatness of wormhole configuration. In order to cure this problem we set $b_3=0$ and find the value of BD scalar field at the throat as
\be\label{bdsfr0}
\phi_0=\f{2(3w-1)r_0^{m+2}}{1+w(m+1)},~~~~~w\neq\left\{-1,\f{1}{3}\right\},~~~\omega\neq-\f{3}{2}.
\ee
With the above choice the condition on asymptotic flatness of wormhole geometry is revised as: $w>1/3$ $\land$ $m<-4$. Also the solution given in Eqs.~(\ref{solb}) and (\ref{solV}) along with the expressions for density and pressure profiles are revised according to the following substitutions
\bea\label{subs}
&&b_2\rightarrow \f{2(1+w)r_0^{-m-4}}{(3w-1)(2\omega+3)},~~~V_2\rightarrow\f{4(1+w)r_0^{m+4}}{1+w(m+1)},~~~V_3\rightarrow0,~~~\rho_1\rightarrow0,~~~\rho_3\rightarrow-\f{r_0^{m+4}}{\pi[1+w(m+1)]},\nn &&p_1\rightarrow0,~~~~p_3\rightarrow \frac{(w+1) (6 \omega +11)r_0^{m+4}}{4 \pi  (2 \omega +3) [1+w(m+1)]}.
\eea
We note that at the throat the conditions $b(r_0)=r_0$ and $\rho(r_0)=\rho_0$ are still applied. Moreover, $b(r)/r\rightarrow0$ and $\rho(r)\rightarrow0$ in the limit $r\rightarrow\infty$. Therefore, the energy conditions Eq.~(\ref{enprpt0}) along with the flare-out condition Eq.~(\ref{flr0}) at the throat are reduced to the following expressions
\bea\label{enconflred}
\rho_0\geq0,~~~~~~~w>-1,~~~~~~~~\rho_0+(\rho_2+p_2)r_0^m+\f{\rho_3+p_3}{r_0^4}\geq0,~~~~~~~~\frac{2 (m+4) (w+1)}{(3 w-1) (2 \omega +3)}<0.
\eea
We therefore observe that the satisfaction of crucial conditions for a physically reasonable wormhole configuration depends on the values of the model parameters $\{\rho_0,w,m,\omega,r_0\}$. However, we need to check whether the tangential profile of NEC in Eq.~(\ref{energypt}) is valid for $r>r_0$. With a more precise investigation we find out that $(3w-1)/w\leq2$ for $1/3<w\leq1$. Since $m<-4$, as the radial coordinate grows, the third and forth terms in this expression tend to zero faster than the first one. Hence we may deduce that if $\rho_0>-\sum_{i=2}^{i=3}(\rho_i+p_i)$ at the throat then the tangential profile of NEC is satisfied for $r>r_0$. Taking into account this condition along with the conditions given in Eq.~(\ref{enconflred}) we arrive at the 2D space parameter constructed by $\{w,\omega\}$ parameters, once we determine the rest ones, i.e., $\{\rho_0,m,r_0\}$. The result is shown in the left panel of Fig.~(\ref{fig1}) where we can find the allowed values of the EoS and BD coupling parameters according to the above mentioned conditions. In the right panel we have plotted the behavior of energy density in terms of radial coordinate and EoS parameter. The BD coupling parameter has been chosen according to the allowed regions given in the left panel. Fig.~(\ref{fig2}) shows the radial and tangential components of NEC against the radial coordinate and EoS parameter where we observe that the NEC is satisfied at $r=r_0$ and $r>r_0$. We therefore conclude that, for the allowed values of model parameters subject to the left panel of Fig.~(\ref{fig1}), both NEC and WEC are fulfilled at the throat and throughout the spacetime. In the left panel of Fig.~(\ref{fig3}) we have sketched the inverse of radial metric component and the right one shows the flare-out condition, where we observe that conditions \ref{c1} and \ref{c2} are satisfied at the throat and for $r>r_0$. Note that for $m=-5$, we have
\be
rb^\prime(r)-b(r)=-\f{2r_0(w+1)}{(3w-1)(2\omega+3)},
\ee
which is negative for $w>1/3$ and $\omega>-3/2$. 
\section{Solutions with $\Phi(r)\neq0$}\label{rednzero}
\begin{figure} 
	\begin{center} 
		\includegraphics[width=8.0cm]{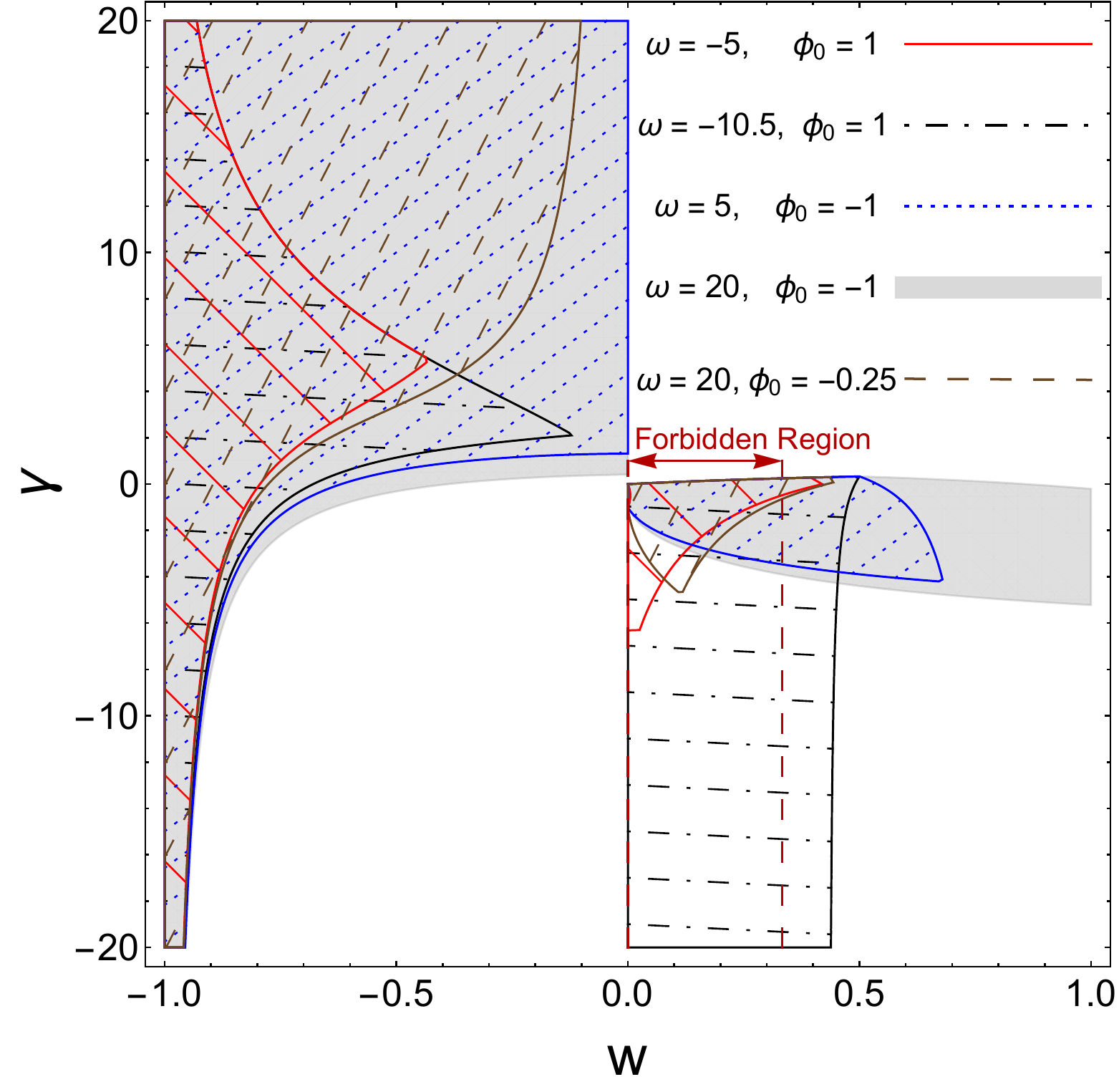}
		\caption{The allowed values of the pair of parameters $(\gamma,w)$ for different values of energy density and BD scalar field at the throat. We have set $\rho_0=1$ and $r_0=1$. Right panel: Energy density versus radial coordinate for the values of model parameters subject to the left panel. We have set $r_0=1$ and $\rho_0=1$.}\label{fig4}
	\end{center}
\end{figure}

\begin{figure}
	\begin{center} 
		\includegraphics[width=7.7cm]{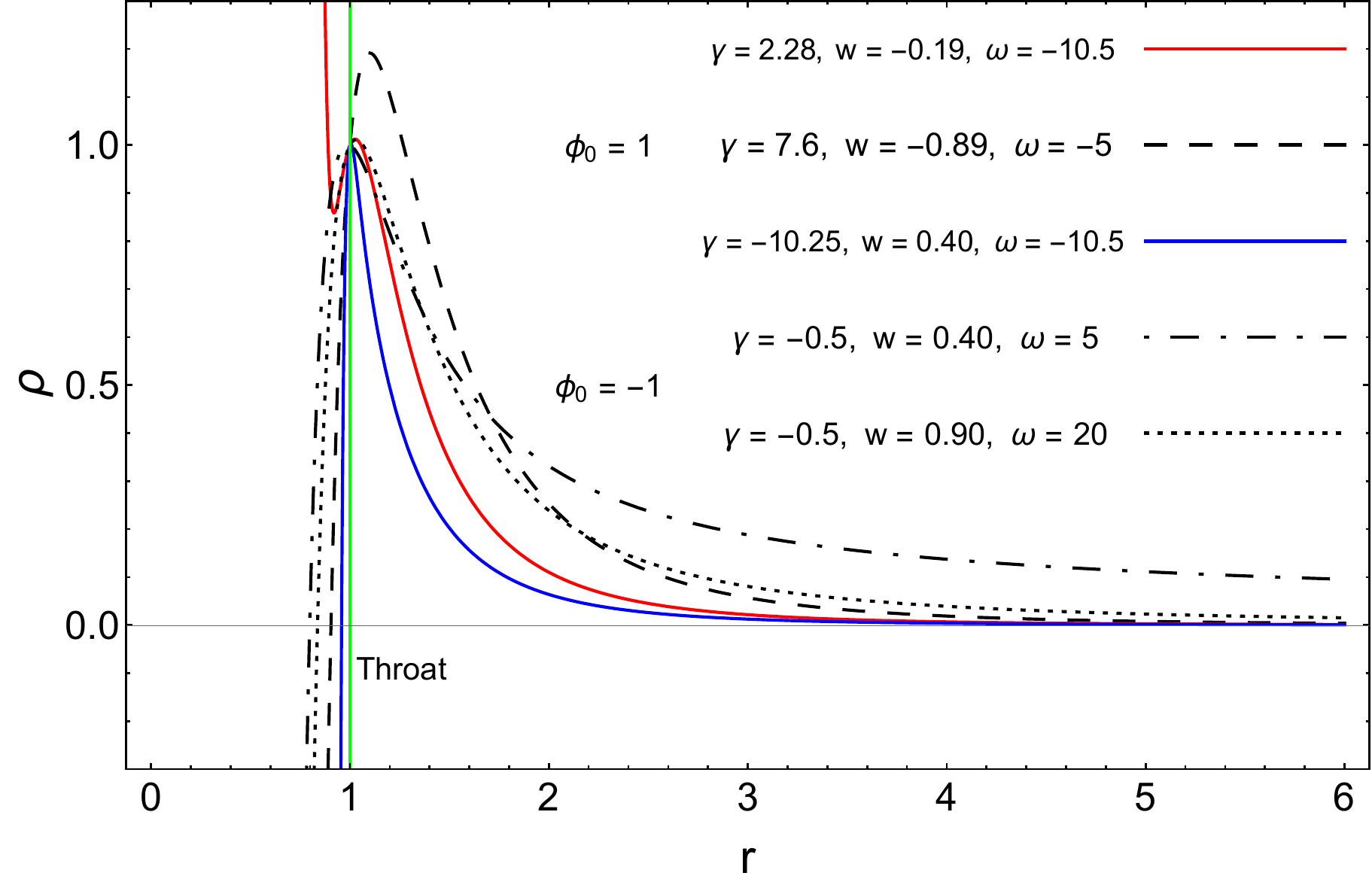}
		\includegraphics[width=7.7cm]{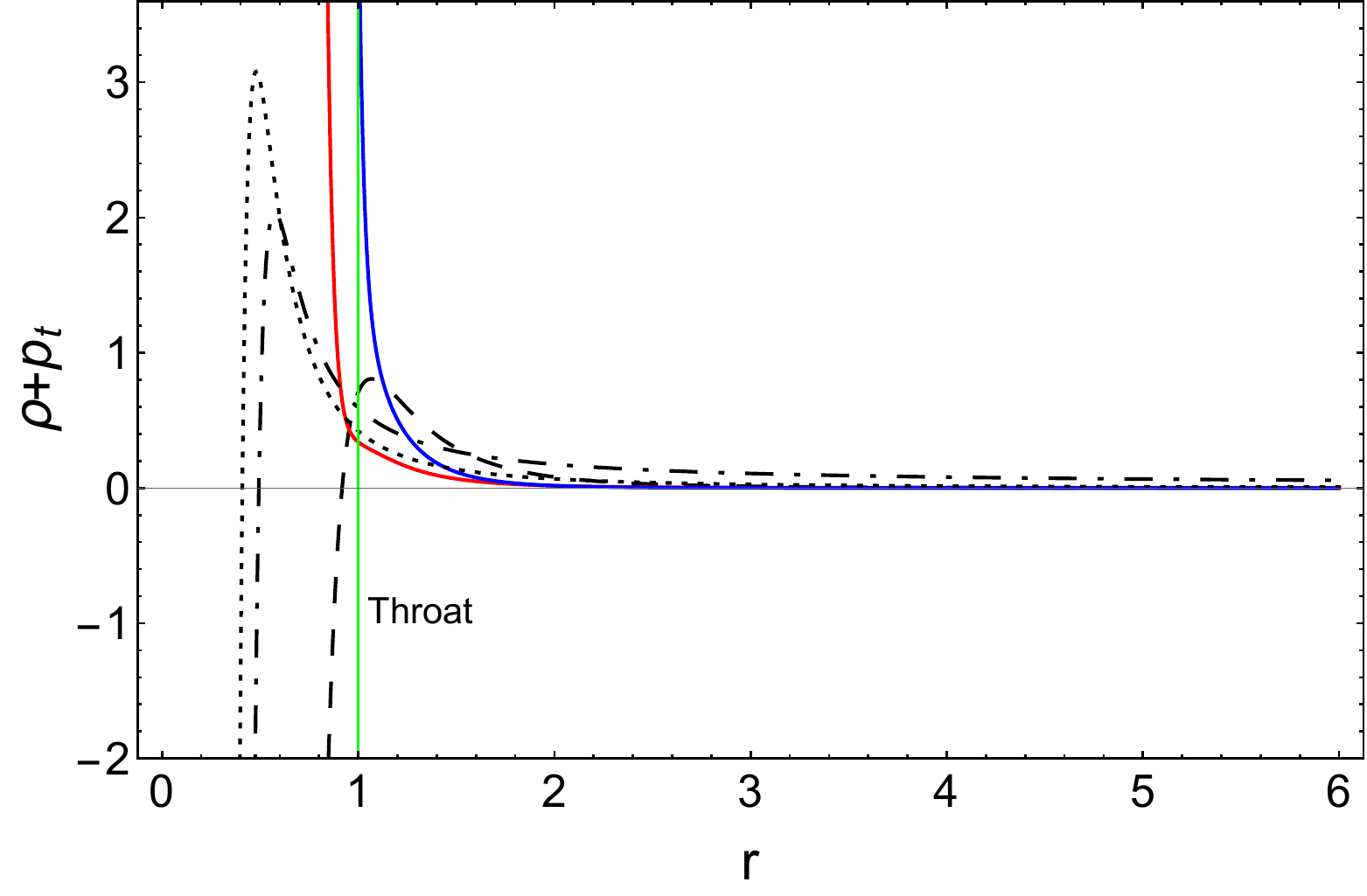}
		\caption{Behavior of energy density (left panel) and NEC in tangential direction (right panel) for allowed values of model parameters according to Fig.~(\ref{fig4}).}\label{fig5}
	\end{center}
\end{figure}

\begin{figure}
	\begin{center} 
		\includegraphics[width=8cm]{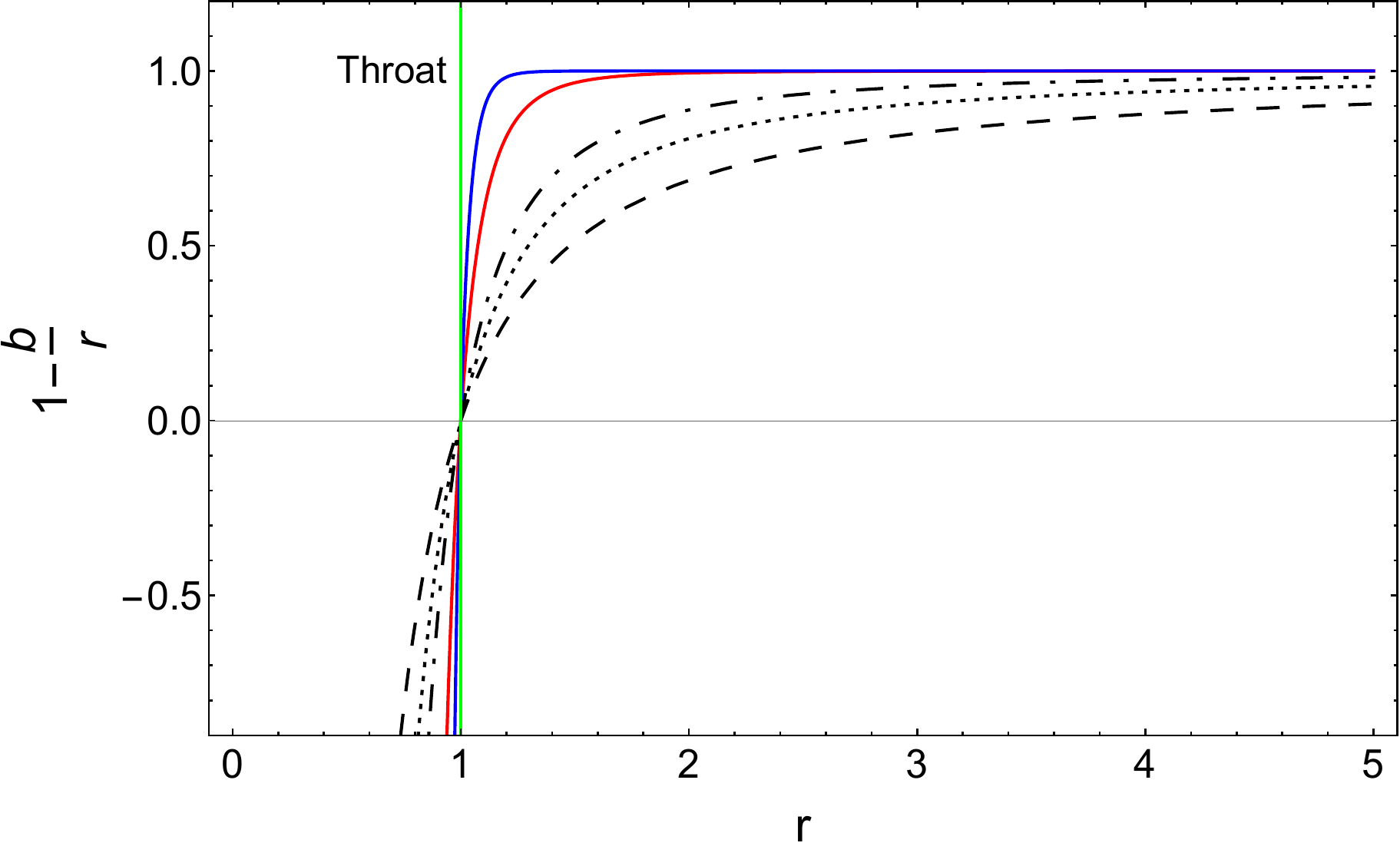}
		\includegraphics[width=7.5cm]{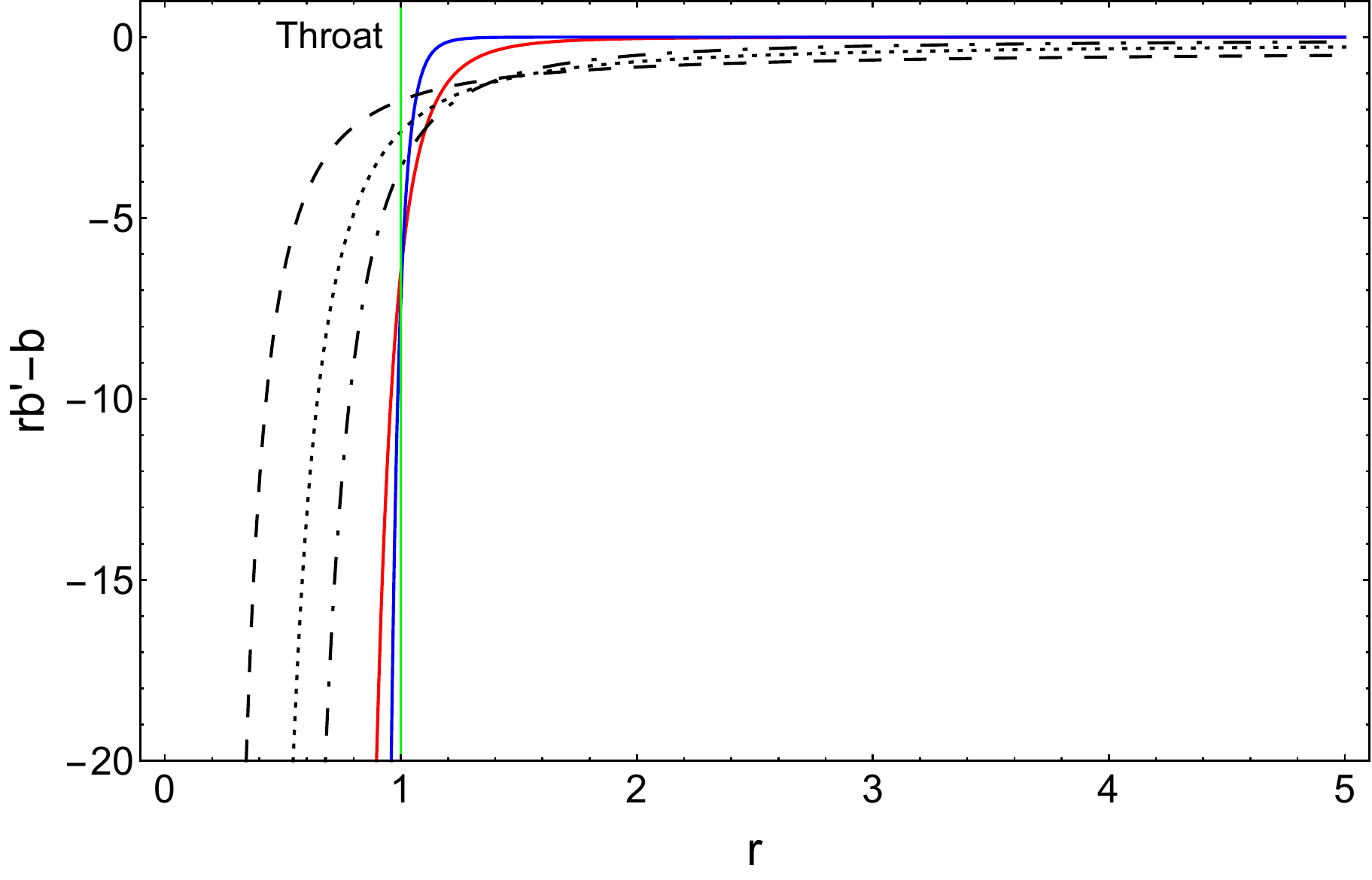}
		\caption{Behavior of $g_{rr}^{-1}$ (left panel) and flare-out condition (right panel) for the same values of model parameters as of Fig.~(\ref{fig5}).}\label{fig6}
	\end{center}
\end{figure}

In the present section we try to find wormhole solutions with nonzero tidal force. {In construction of wormhole solutions, a variable redshift function offers key advantages, including nontrivial wormhole spacetimes satisfying NEC without the presence of event horizons (ensuring traversability), making control over tidal forces, and compatibility with asymptotic flatness. By adjusting the redshift function, researchers can fine-tune gravitational redshift effects and minimize stress-energy requirements for exotic matter, making wormhole solutions more physically plausible, see~\cite{khu},\cite{mt},\cite{newast2024},\cite{nonzeroredsh} for further details.} We therefore proceed to solve Eqs.~(\ref{bdrhowpr}) and (\ref{bdtovrhosol}) taking the following assumptions: $i)$ the functionality of BD scalar field is given by, $\phi(r)=\phi_0\left(\f{r_0}{r}\right)^2$, $ii)$ the energy density is given by Eq.~(\ref{rhosol}), $iii)$ the redshift function is defined as
\be\label{redshiftf}
\Phi(r)=\Phi_0+\gamma\left(\f{1}{r}-\f{1}{r_0}\right),
\ee 
where $\Phi_0$ is the value of redshift function at the throat and $\gamma$ is a constant parameter. Equations (\ref{bdrhowpr}) and (\ref{bdtovrhosol}) then provide us with the following solution 
\bea
\Phi(r)\!\!\!&=&\!\!\!\int[r-b(r)]^{-1}\left[\f{wb^\prime(r)}{(w+1)}+\f{(1-4w)b(r)}{r(w+1)}-\f{4r^4f(r)}{\phi_0r_0^2(2\omega+3)}+\f{w(6\omega+7)-2\omega-5}{(w+1)(2\omega+3)}\right]dr,\label{phisolnonz}\nn\\
V(r)\!\!\!&=&\!\!\!\f{2\phi_0r_0^2}{r^5(w+1)}\Big\{(2\omega+3)(w-1)b(r)-2r\left[\omega(w-1)+w-2\right]\Big\}.\label{Vsolnonz}
\eea
Now, we set the integrand within Eq.~(\ref{phisolnonz}) equal to first derivative of the redshift function and solve the resultant differential equation for the shape function, this gives
\bea
b(r)=\Bigg\{\int g(r)r^{\f{1}{w}-5}{\rm exp}\left[\f{\gamma(w+1)}{wr}\right]dr+{\rm C}_3\Bigg\}r^{4-\f{1}{w}}{\rm exp}\left[{-\f{\gamma(w+1)}{wr}}\right],~~~~~~~~w\neq0,\label{bsuggest}
\eea
where 
\be\label{gfunction}
g(r)=\f{4r^5(w+1)f(r)-\phi_0r_0^2\Big[r\left[(6\omega+7)w-2\omega-5\right]+\gamma(2\omega+3)(w+1)\Big]}{w\phi_0r_0^2(2\omega+3)},~~~~~~\omega\neq-\f{3}{2},
\ee
and ${\rm C}_3$ is an integration constant. In order to find a reasonable solution for the shape function one may consider different forms for the auxiliary function $f(r)$. Let us consider the case for which
\bea\label{auxfr}
g(r)={\rm C}_0{\rm exp}\left[-\f{\gamma(1+w)}{wr}\right]~~~\!\!\!\!\!\!\!\!\!\!\!\!\!\!&&\Rightarrow\!\!\!\!~~~f(r)=\f{\phi_0r_0^2w(2\omega+3)}{4(w+1)r^5}\times\nn&&\!\!\!\!\!\!\!\!\!\!\!\!\!\!\!\!\!\Bigg[{\rm C}_0{\rm exp}\left[-\f{\gamma(1+w)}{wr}\right]+\f{(2\omega+3)[r(3w-1)+\gamma(1+w)]-2r(w+1)}{w(2\omega+3)}\Bigg],
\eea
where ${\rm C}_0$ is a constant. With the above assumption, the shape function is found as
\be\label{shapefin}
b(r)=r_0{\rm exp}\left[\f{\gamma(w+1)}{w}\left(\f{1}{r_0}-\f{1}{r}\right)\right],~~~~~{\rm C}_0=\f{r_0(1-4w)}{w}{\rm exp}\left[\f{\gamma(1+w)}{wr_0}\right],
\ee
where we have set the constant ${\rm C}_3=0$ so that $b(r_0)=r_0$. It is a straightforward task to check that substituting the above solution along with the assumption Eq.~(\ref{auxfr}) into Eq.~(\ref{phisolnonz}), the redshift function Eq.~(\ref{redshiftf}) is recovered. Regarding the above solution we find that condition~\ref{c1} is satisfied at the throat. The flare-out condition at the throat gives the inequality $(1+w)\gamma/wr_0<1$. Condition~\ref{c3} implies that
\be\label{cond3mp}
\f{b(r)}{r}\Bigg|_{r\rightarrow\infty}\!\!\!\!\!\approx~r_0{\rm exp}\left[\f{\gamma(1+w)}{wr_0}\right]\left(\f{1}{r}-\f{\gamma(1+w)}{wr^2}+\f{(1+w)^2\gamma^2}{2w^2r^3}+{\mathcal O}\left[\f{1}{r}\right]^4\right),
\ee
whence we readily find that $b(r)/r\rightarrow 0$ at spatial infinity. Also, the condition~\ref{c4} follows the discussions given in the previous section. Substituting Eq.~(\ref{shapefin}) into Eq.~(\ref{Vsolnonz}) gives the following form for the scalar field potential
\bea\label{sfpot}
V(r)=\f{2r_0^2\phi_0}{(w+1)r^5}\left\{r_0(w-1)(2\omega+3){\rm exp}\left[\f{\gamma(w+1)(r-r_0)}{wrr_0}\right]-2r\left[\omega(w-1)+w-2\right]\right\}.
\eea
The scalar field potential assumes the value $V_0=2\phi_0/r_0^2$ at the throat. Now, we can find the energy density by solving the differential equation Eq.~(\ref{diffeqrhof}) or the integral given in Eq.~(\ref{rhosol}). We therefore substitute for the redshift function Eq.~(\ref{redshiftf}), the scalar field potential Eq.~(\ref{sfpot}) and the auxiliary function Eq.~(\ref{auxfr}) into Eq.~(\ref{rhosol}) to find the energy density as
\bea\label{rhosolnon}
\rho(r)&=&\f{r_0^2}{4\pi r^5(w+1)}\Bigg\{r_0{\rm exp}\left[\f{\gamma(w+1)(r-r_0)}{wrr_0}\right]\left(4\pi\rho_0 r_0^2\left(\f{r}{r_0}\right)^{2+\f{1}{w}}(1+w)+\phi_0(2\omega+3)\right)\nn&-&\phi_0(2\omega+3)r\Bigg\},
\eea
where the integration constant has been set in such a way that $\rho(r_0)=\rho_0$. From condition \ref{c3} we require that the EMT components vanish at spatial infinity. This gives
\bea\label{rhosolnonass}
\rho(r)\Big|_{r\rightarrow\infty}\!\!\!\approx~r^{\f{1}{w}}\left[\f{\rho_0}{r^3}{\rm exp}\left[\f{\gamma(1+w)}{wr_0}\right]r_0^{3-\f{1}{w}}+{\mathcal O}\left(\f{1}{r}\right)^4\right]+\left[-\f{\phi_0r_0^2(2\omega+3)}{4\pi(w+1)r^4}+{\mathcal O}\left(\f{1}{r}\right)^5\right],
\eea
whence we can figure out that the energy density vanishes asymptotically, if
\be\label{assen}
-1\leq w<0~~~\lor~~~w>1/3,
\ee
where the lower bound is imposed from the satisfaction of radial profile of WEC. According to the equation of state, we find that the asymptotic behavior of radial pressure follows that of energy density. The tangential pressure is also obtained as
\bea\label{tanpressnon}
p_t(r)&=&\f{\phi_0r_0^2}{16\pi w(w+1)r^7}\left[2wr\xi_1(r)-r_0{\rm exp}\left[\f{\gamma(w+1)(r-r_0)}{wrr_0}\right]\xi_2(r)\right],
\eea
where
\bea\label{coeffpt}
\xi_1(r)\!\!\!&=&\!\!\!2 r^2 (w+2 \omega +4)+3 \gamma  r (w+1)+\gamma ^2 (w+1),\nn
\xi_2(r)\!\!\!&=&\!\!\!r^2 w (3 w+8 \omega +15)+\gamma  r (w+1) (6 w-1)+\gamma ^2 \left(w^2-1\right).
\eea
From the above expression we get the asymptotic behavior of tangential pressure as
\be\label{tanpressass}
p_t(r)\Big|_{r\rightarrow\infty}\!\!\!\approx~\f{\phi_0r_0^2(2\omega+w+4)}{4\pi(w+1)r^4}+{\mathcal O}\left(\f{1}{r}\right)^5,
\ee
from which we readily find that the tangential pressure vanishes at spatial infinity. From Eqs.~(\ref{rhosolnon}) and (\ref{tanpressnon}) we arrive at following expressions for radial and tangential profiles of NEC 
\bea
\rho(r)+p_{t}(r)\!\!\!&=&\!\!\!\f{\phi_0r_0^2}{8\pi r^6}(r+\gamma)(2r+\gamma)+{\rm exp}\left[\f{\gamma(w+1)(r-r_0)}{wrr_0}\right]\left[\rho_0\left(\f{r_0}{r}\right)^{3-\f{1}{w}}-\f{\phi_0r_0^3}{16w\pi r^7}\xi_3(r)\right]\geq0,\nn\rho(r)+p_{r}(r)\!\!\!&=&\!\!\!(1+w)p_r(r)\geq0,~~~\xi_3(r)=3 r^2 w+\gamma  r (6 w-1)+\gamma ^2 (w-1).\label{wecexp}
\eea
The WEC at the throat leads to the following inequalities
\bea
\rho\Big|_{r=r_0}\!\!\!\!\!&=&\rho_0\geq0,~~~~~~~~\rho+p_{r}\Big|_{r=r_0}\!\!=(1+w)\rho_0\geq0,~~~~~~~~\nn\rho+p_{t}\Big|_{r=r_0}\!\!\!\!&=&\rho_0+\f{\phi_0\left[r_0^2 w+\gamma  r_0+\gamma ^2 (w+1)\right]}{16 \pi  r_0^4 w}\geq0.\label{enprptnon}
\eea
Now, we are at a situation to find possible bounds on model parameters according to conditions \ref{c2} and \ref{c3} along with WEC and NEC. We further require that the slope of the energy density curve be positive at the throat. This helps us to have positive values for energy density beyond the throat. In this sense, the energy density can obey an increasing function of radial coordinate with $\rho(r_0)>0~\land~\rho^\prime(r_0)>0$. Since the energy density curve has to vanish asymptotically, it must then reaches a maximum value for $r>r_0$ after which $\rho(r)\rightarrow0$ as $r\rightarrow\infty$. It is still possible to consider the case with $\rho^\prime(r_0)<0$, for which, the energy density curve can assume a decreasing function of radial coordinate subject to the conditions $\rho(r_0)>0$ and $\rho(r)\rightarrow0$ at spatial infinity. Let us consider the former case for which we require that
\bea\label{drhodrcond}
\f{d\rho(r)}{dr}\Bigg|_{r=r_0}\!\!\!\!\!\!\!=~\f{\rho_0}{wr_0^2}\left[r_0-3r_0w+\gamma(1+w)\right]-\f{\phi_0(2\omega+3)\left[wr_0-\gamma(1+w)\right]}{4\pi r_0^4w(1+w)}>0.
\eea
Regarding the above mentioned conditions we find the following restrictions on model parameters
\bea\label{resticphip}
&&r_0>0\land\rho_0>0\land\Bigg[\left(-1<w<0\land\f{wr_0}{1+w}<\gamma\leq-\f{r_0}{2(w+1)}+\f{\alpha_1}{2}\land\omega<\alpha_2\right)\lor\nn&&\left(\f{1}{3}<w\leq1\land\gamma<\f{wr_0}{1+w}\land\omega<\alpha_2\right)\Bigg],
\eea
for $\phi_0>0$ and 
\bea\label{resticphin}
\!\!\!\!\!\!\!\!\!\!&&r_0>0\land \phi_0\beta_1[r_0 w-\gamma  (w+1)]>0\land\Bigg\{\Bigg[8 \pi  \rho_0 r_0+\frac{\phi_0}{r_0}=0\land \Bigg[\Bigg(w+\beta_2 +\frac{1}{2}=0\land\nn\!\!\!\!\!\!\!\!\!\!&& (w+1) \left[r_0+(w+1) \left(\alpha_1^{\f{1}{2}} +2 \gamma \right)\right]>0\Bigg)\lor \left(w+1>0\land \gamma +\frac{r_0}{2 w+2}\geq \frac{\alpha_1^{\f{1}{2}}}{2}\land w+\beta_2 +\frac{1}{2}<0\right)\lor\nn\!\!\!\!\!\!\!\!\!\!&& \left[w+\beta_2 +\frac{1}{2}>0\land w<0\land (w+1) \left[(w+1)\left(\alpha_1^{\f{1}{2}}-2\gamma \right)-r_0\right]<0\right]\lor \Bigg(3 w>1\land\nn\!\!\!\!\!\!\!\!\!\!&& (w+1) \left[(w+1) \left(\alpha_1^{\f{1}{2}}-2\gamma\right)-r_0\right]>0\land \beta_3 \geq 0\land w\leq 1\Bigg)\Bigg]\Bigg]\lor\Bigg[256 \pi  \rho_0+\frac{7 \phi_0}{r_0^2}>0\land 128 \pi  \rho_0+\frac{7 \phi_0}{r_0^2}\leq 0\land\nn\!\!\!\!\!\!\!\!\!\!&& \Bigg[\left(w+\frac{1}{2}=\beta_2 \land \beta_3 =0\right)\lor\left(w+1>0\land \gamma +\frac{r_0}{2 w+2}\geq \frac{\alpha_1^{\f{1}{2}}}{2}\land w<0\right)\lor \Big(3 w>1\land \beta_3 \geq 0\land w+\frac{1}{2}<\beta_2 \nn\!\!\!\!\!\!\!\!\!\!&& \land\gamma +\frac{r_0}{2 w+2}\leq \frac{\alpha_1^{\f{1}{2}} }{2}\Big)\Bigg]\Bigg]\lor\Bigg[8 \pi  \rho_0 r_0^4+r_0^2 \phi_0>0\land \Bigg[\left(w<0\land w+\beta_2 +\frac{1}{2}\geq 0\land \gamma >\frac{r_0 w}{w+1}\right)\lor\nn\!\!\!\!\!\!\!\!\!\!&& \Bigg[w+1>0\land w+\beta_2 +\frac{1}{2}<0\land\left[\left(\frac{r_0 w}{w+1}<\gamma \land \beta_3 \leq 0\right)\lor \gamma +\frac{r_0}{2 w+2}\geq \frac{\alpha_1^{\f{1}{2}}}{2}\right]\Bigg]\lor\nn\!\!\!\!\!\!\!\!\!\!&& \Bigg(\gamma <\frac{r_0 w}{w+1}\land \beta_3 \geq 0\land w\leq 1\land 3 w>1\Bigg)\Bigg]\Bigg]\lor \Bigg[w+1>0\land \gamma +\frac{r_0}{2 w+2}\geq \frac{\alpha_1^{\f{1}{2}}}{2}\land w<0\land\nn\!\!\!\!\!\!\!\!\!\!&& \Bigg[\left(128 \pi  \rho_0+\frac{7 \phi_0}{r_0^2}>0\land 8 \pi  \rho_0+\frac{\phi_0}{r_0^2}<0\right)\lor \Big(\rho_0>0\land 256 \pi  \rho_0+\frac{7 \phi_0}{r_0^2}\leq 0\Big)\Bigg]\Bigg]\lor \nn\!\!\!\!\!\!\!\!\!\!&&\Bigg[128 \pi  \rho_0+\frac{7 \phi_0}{r_0^2}>0\land 3 w>1\land \beta_3 \geq 0\land 8 \pi  \rho_0+\frac{\phi_0}{r_0^2}<0\land w\leq 1\land \gamma +\frac{r_0}{2 w+2}\leq \frac{\alpha_1^{\f{1}{2}}}{2}\Bigg]\Bigg\},
\eea
for $\phi_0<0$, where
\bea\label{alphabetas}
\alpha_1&=&\f{\phi_0r_0^2[1-4w(w+1)]-64\pi\rho_0r_0^4w(w+1)}{\phi_0(w+1)^2},\nn
\alpha_2&=&-\f{3}{2}-\f{2\pi\rho_0r_0^2(w+1)\left[r_0(3w-1)-\gamma(1+w)\right]}{\phi_0[wr_0-\gamma(1+w)]},\nn
\beta_1&=&4 \pi  \rho_0 r_0^2 (w+1) [r_0 (3 w-1)-\gamma  (w+1)]+(2 \omega +3) \phi_0[r_0 w-\gamma  (w+1)],\nn
\beta_2&=&\left[\f{8\pi r_0^2\rho_0+\phi_0}{32\pi r_0^2\rho_0+2\phi_0}\right]^{\f{1}{2}},~~~~~~~~~~~~~\beta_3=\f{r_0+(w+1)(\alpha_1^{\f{1}{2}}+2\gamma)}{w+1}.
\eea
The inequalities Eqs.~(\ref{resticphip}) and (\ref{resticphin}) are general bounds on model parameters $\{\phi_0,\rho_0,r_0,\gamma,w,\omega\}$. These set of parameters construct a six-dimensional space parameter that the allowed regions, subject to the aforementioned conditions, present physically reasonable wormhole solutions. As this space parameter cannot be sketched in a usual way we consider a two-dimensional subspace of it by determining the values of the parameters $\{\phi_0,\rho_0,r_0,\omega\}$. The result is shown in Fig.~(\ref{fig4}) where we have plotted the allowed regions of a two-dimensional subspace constructed out of the pair of parameters $(\gamma,w)$. These regions have been provided for special values of the throat radius and values of energy density and BD scalar field at the throat. The two vertical dashed lines show a forbidden region for which $0<w<1/3$. For these values of EoS parameter we have $\rho(r)\rightarrow\infty$ as $r\rightarrow\infty$ which is in contrast to asymptotic flatness condition. The left panel in Fig.~(\ref{fig5}) shows the behavior of energy density for different allowed values of model parameters, where we observe that $\rho(r)>0$ for $r\geq r_0$. The right panel in this figure presents the tangential profile of NEC where it is seen that this condition is fulfilled at the throat and beyond it. The behavior of $g_{rr}^{-1}$ for allowed values of the pair of parameters $(\gamma,w)$ is plotted in the left panel of Fig.~(\ref{fig6}) where, we observe that conditions \ref{c1}, \ref{c3} and \ref{c4} are satisfied; also flare-out condition is plotted in the right panel where it is seen that condition \ref{c2} is fulfilled. 
\section{Some further issues}\label{sofuis}
So far, we have obtained a class of exact wormhole solutions for which the supporting matter respects the WEC and NEC along with the physical conditions governing the wormhole geometry. Hence, these configurations do not require the use of exotic matter to construct them. In the following we try to explore some properties of the obtained solutions. 
\subsection{Special cases}
The spacetime representing the Schwarzschild solution can be interpreted as an asymptotically flat wormhole solution, for which the shape function reads $b(r)=r_0={\rm constant}$. Such a wormhole configuration is not traversable due to the existence of a horizon at its throat. Nevertheless, Morris, Thorne and Yurtsever have shown that Schwarzschild traversable wormhole can be constructed by demanding that the redshift function does not admit horizons and the radial metric component assumes the form $g_{rr}^{-1}=1-r_0/r$~\cite{mt,mt1}. Recently these type of wormhole solutions have been studied in detail in~\cite{catal2017},\cite{niltonuni},\cite{casschwa} and in the present work we can recover such wormhole configurations by setting $m=-5$. Then, from Eq.~(\ref{solb}) we get
\be\label{schww}
ds^2=-dt^2+\f{dr^2}{(1-\beta)\left[1-\f{r_0}{r}\right]}+r^2d\Omega^2,
\ee
where
\be\label{betaeq}
\beta(w,\omega)=1-\f{2(1+w)}{(3w-1)(2\omega+3)},~~~~~~~~~~~w\neq\f{1}{3},~~~~~\omega\neq-\f{3}{2}.
\ee
For $\beta<1$, the spacetime metric Eq.~(\ref{schww}) represents a zero tidal force Schwarzschild-like wormhole~\cite{catal2017}. The case with $\beta=0$ describes Schwarzschild wormhole for which the BD parameter assumes the form $\omega(w)=(5-7w)/2(3w-1)$. The blue dashed curve in the left panel of Fig.~(\ref{fig1}) shows the behavior of BD parameter as a function of EoS parameter, where we observe that, all the points in $\left(w,\omega\right)$ plane which lie on this curve correspond to Schwarzschild wormhole solution. Those beyond this curve that are located in the regions with $m=-5$ correspond to Schwarzschild-like wormhole solutions. It should be noted that, using the new radial coordinate $d\tilde{r}=(1-\beta)dr$, the spacetime metric Eq.~(\ref{schww}) can be recast into the following form
\be\label{spmetrec}
ds^2=-dt^2+\f{d\tilde{r}^2}{1-\f{\tilde{r}_0}{\tilde{r}}}+(1-\beta)\tilde{r}^2d\Omega^2.
\ee
The above line element has a solid angle deficit for $0<\beta<1$ and solid angle excess for $\beta<1$. We further note that solid angle of a sphere of unit radius is $4\pi(1-\beta)<4\pi$ for $0<\beta<1$ and $4\pi(1-\beta)>4\pi$ for $\beta<0$. Another type of solutions can be obtained for $m=-6$ for which the spacetime metric reads
\be
ds^2=-dt^2+\f{d\tilde{r}^2}{1-\f{\tilde{r}_0^2}{\tilde{r}^2}}+(1-\beta)\tilde{r}^2d\Omega^2.
\ee
The above metric represents the famous MT spacetime~\cite{mt} with same arguments on solid angle deficit/excess as those of the Schwarzschild wormhole. It should be noted that, regarding the left panel of Fig.~(\ref{fig1}), these class of wormhole solutions hold for points in $\left(\omega,w\right)$ plane which lie on the blue dashed curve and within the black dotted regions. Also for solutions presented in Sec.~(\ref{rednzero}), the special case in which $\gamma=0$ corresponds to a zero redshift Schwarzschild solution Eq.~(\ref{schww}) with $\beta=0$. One may also compare the bounds on BD parameter obtained in the present work with those reported in the previous studies. To do so, we may set the model parameters in such a way that the BD coupling parameter falls within the range $-2<\omega<-4/3$. This interval has been obtained in~\cite{bd3},\cite{bd3rev} where static spherically symmetric wormhole solutions in vacuum BD theory were investigated. For this interval we can find the following inequalities for e.g., $1/3<w\leq1$, as 
\bea\label{newineqs}
\!\!\!\!\!\!\!\!\!\!\!\!\!\!\!\!\!\!\!\!&&\Bigg\{\!\!-2<\omega <-\frac{3}{2}\land \Bigg[\Bigg(\gamma>0\land\rho_0>0\land \Big(\left[\gamma(1+w)+1=3 w\land 3 \gamma <1\right]\lor \Big(\frac{\gamma +1}{3-\gamma }<w\land\rho_0+\epsilon_0<0\nn\!\!\!\!\!\!\!\!\!\!\!\!\!\!\!\!\!\!\!\!&&\land w\leq 1\land 4 \gamma \leq 1\Big)\Big)\!\Bigg)\!\lor\!\Bigg[\rho_0>0\land \rho_0+\epsilon_0<0\land w\leq 1\land \Bigg(\!\!\left(4 \gamma >1\land 3 \gamma <1\land \frac{\gamma +1}{3-\gamma }<w\right)\lor\nn\!\!\!\!\!\!\!\!\!\!\!\!\!\!\!\!\!\!\!\!&& \left(3 w>1\land \gamma \leq 0\right)\!\lor\!\left(2 \gamma <1\land \frac{1}{\gamma -1}+w+1>0\land 3 \gamma \geq1\right)\!\!\Bigg)\Bigg]\Bigg]\Bigg\}\!\lor\!\Bigg\{\!\Bigg[\!\left(4 \gamma \leq 1\land \frac{1}{3}<w<\frac{\gamma +1}{3-\gamma }\right)\nn\!\!\!\!\!\!\!\!\!\!\!\!\!\!\!\!\!\!\!\!&&\lor\left(\frac{1}{4}<\gamma <\frac{1}{3}\land \gamma +\gamma  w<w\land \gamma +\gamma  w+1>3 w\right)\Bigg]\land\Bigg[\left(\rho_0>0\land -2<\omega < -\frac{3}{2}\right)\lor\nn\!\!\!\!\!\!\!\!\!\!\!\!\!\!\!\!\!\!\!\!&& \left(\rho_0>-\epsilon_0\land -\frac{3}{2}<\omega <-\frac{4}{3}\right)\Bigg]\Bigg\},
\eea
where
\bea\label{epsil0}  
&&\epsilon_0=\frac{(2 \omega +3) (\gamma +(\gamma -1) w)}{4 \pi  (w+1) (\gamma +(\gamma -3) w+1)},
\eea
and we have set $r_0=\phi_0=1$. Also, in~\cite{bd4}, exact wormhole solutions in vacuum BD theory were obtained for which the BD coupling parameter lies in the range $\omega<-2~\lor~-2<\omega\leq0$. For these values of coupling parameter we find the following bounds on model parameters 
\bea\label{omeg2}
&&\Bigg\{\rho_0>0\land \Bigg[\Bigg[3 \gamma <1\land \Bigg(\!\!\left(\gamma >0\land \frac{4}{\gamma -3}+w+1=0\land 2 \omega +3<0\land \omega +2\neq 0\right)\lor\nn&& \left(4 \gamma >1\land \frac{1}{\gamma -1}+w+1>0\land \frac{4}{\gamma -3}+w+1<0\land \left(\omega <-2\lor -2<\omega < -\frac{3}{2}\right)\right)\!\!\Bigg)\Bigg]\lor\nn&& \left(3 w>1\land 4 \gamma \leq 1\land \frac{4}{\gamma -3}+w+1<0\land \left(\omega <-2\lor -2<\omega <-\frac{3}{2}\right)\!\right)\!\Bigg]\!\Bigg\}\lor\nn&& \Bigg\{\rho_0>0\land \rho_0+\epsilon_0<0\land w\leq 1\land \Bigg[\left(0<\gamma <\frac{1}{3}\land \gamma +\gamma  w+1<3 w\right)\lor\nn&& \left(\frac{1}{\gamma -1}+w+1>0\land \frac{1}{3}\leq \gamma <\frac{1}{2}\right)\lor (3 w>1\land \gamma \leq 0)\Bigg]\land \left(\omega <-2\lor -2<\omega <-\frac{3}{2}\right)\!\!\Bigg\}\lor\nn&& \Bigg[\rho_0+\epsilon_0>0\land 2 \omega +3>0\land \frac{4}{\gamma -3}+w+1<0\land \omega \leq 0\land \Bigg[(\gamma >0\land 3 w>1\land 4 \gamma \leq 1)\lor\nn&& \left(4 \gamma >1\land 3 \gamma <1\land \frac{1}{\gamma -1}+w+1>0\right)\!\!\Bigg]\Bigg],
\eea
where we have considered the same conditions on $\{r_0,\phi_0,w\}$ as those for which Eq.~(\ref{newineqs}) was obtained.
\subsection{Proper distance and embedding diagrams}
\begin{figure}
	\begin{center} 
		\includegraphics[width=8cm]{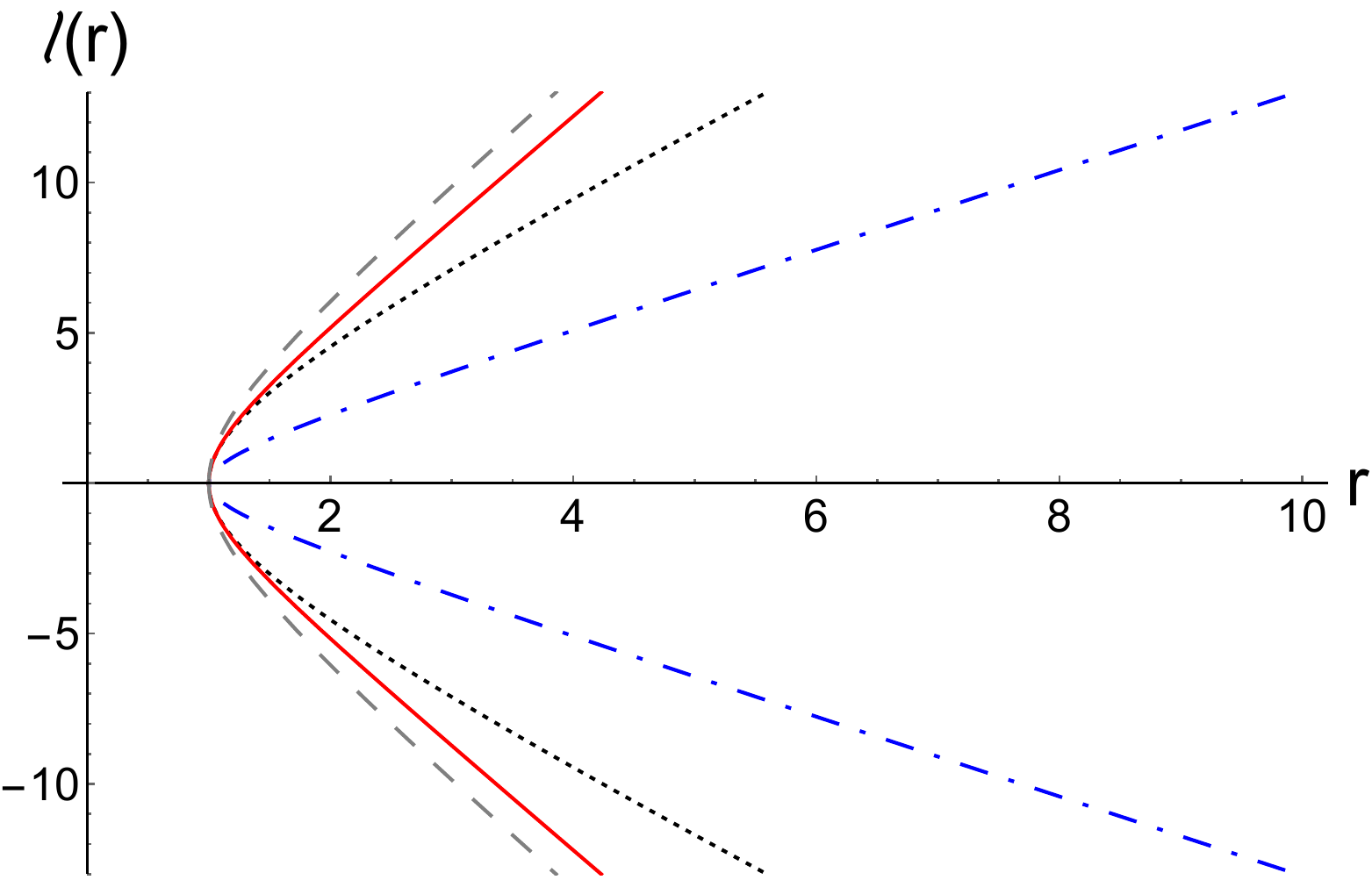}
		\includegraphics[width=7.5cm]{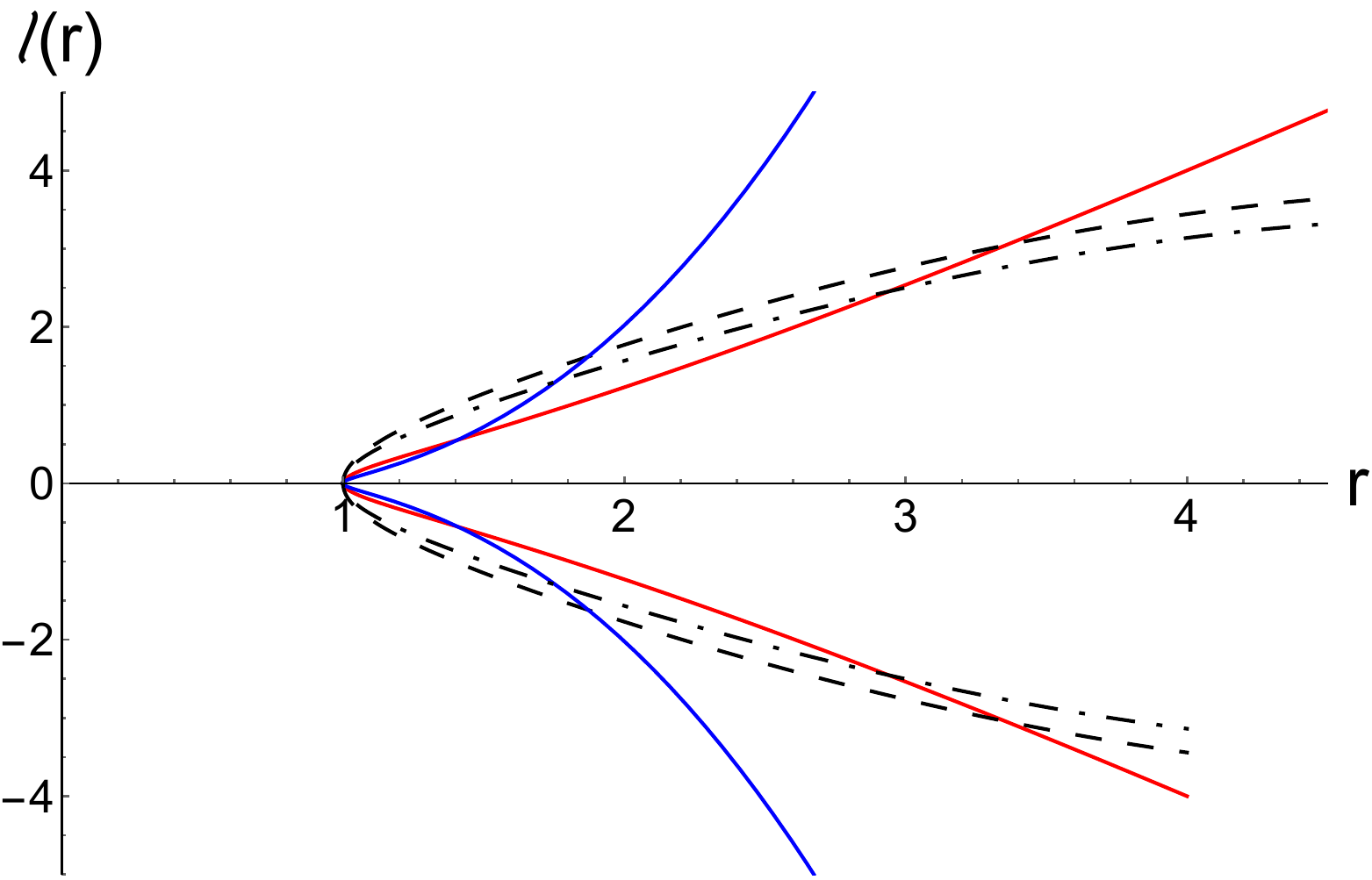}
		\caption{Proper distance against radial coordinate for zero (left panel) and nonzero (right panel) redshift solutions. The model parameters have been set according to Figs.~(\ref{fig3}) and~(\ref{fig5}).}\label{fig7}
	\end{center}
\end{figure}

\begin{figure}
	\begin{center} 
		\includegraphics[width=8cm]{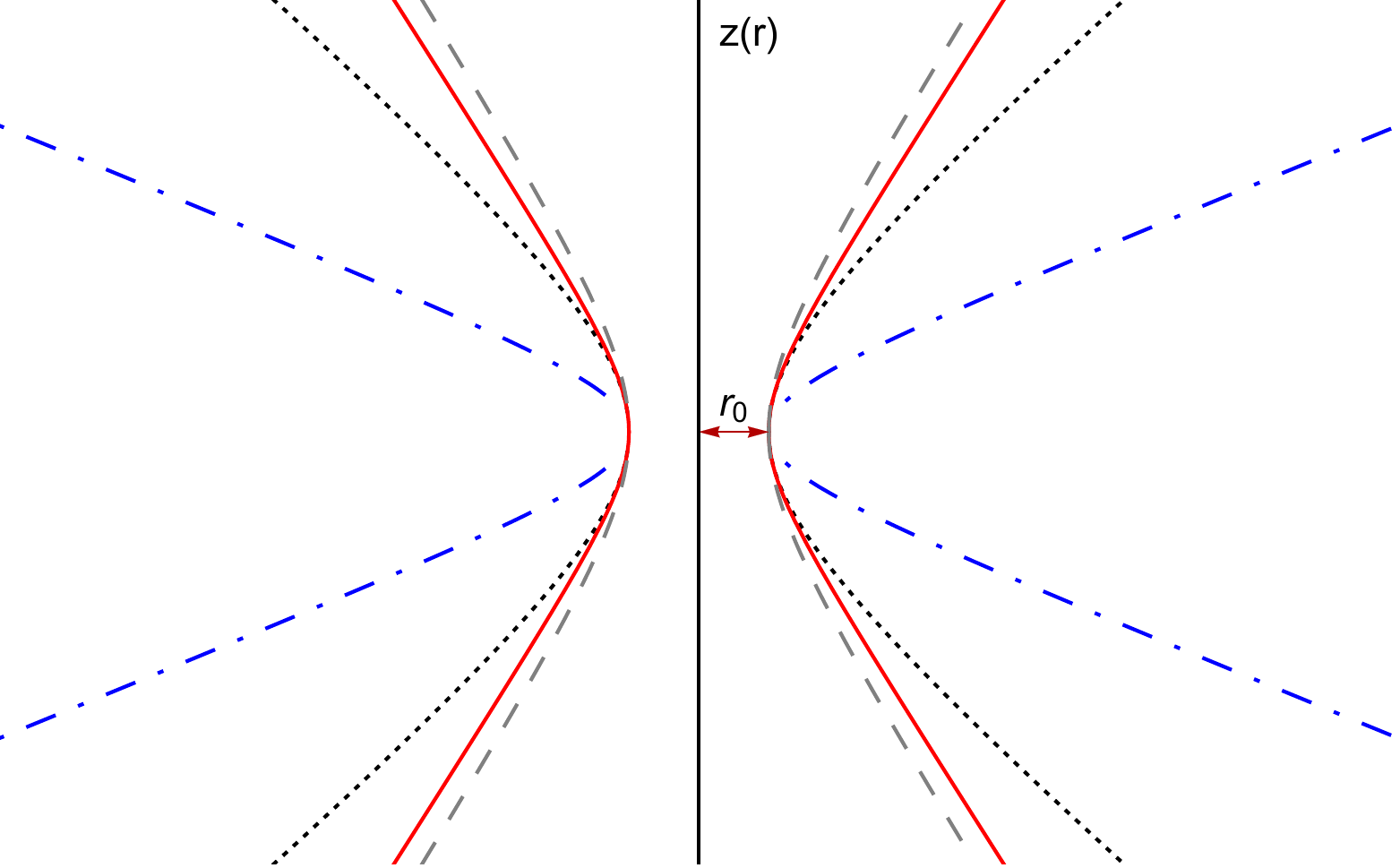}
		\includegraphics[width=7.5cm]{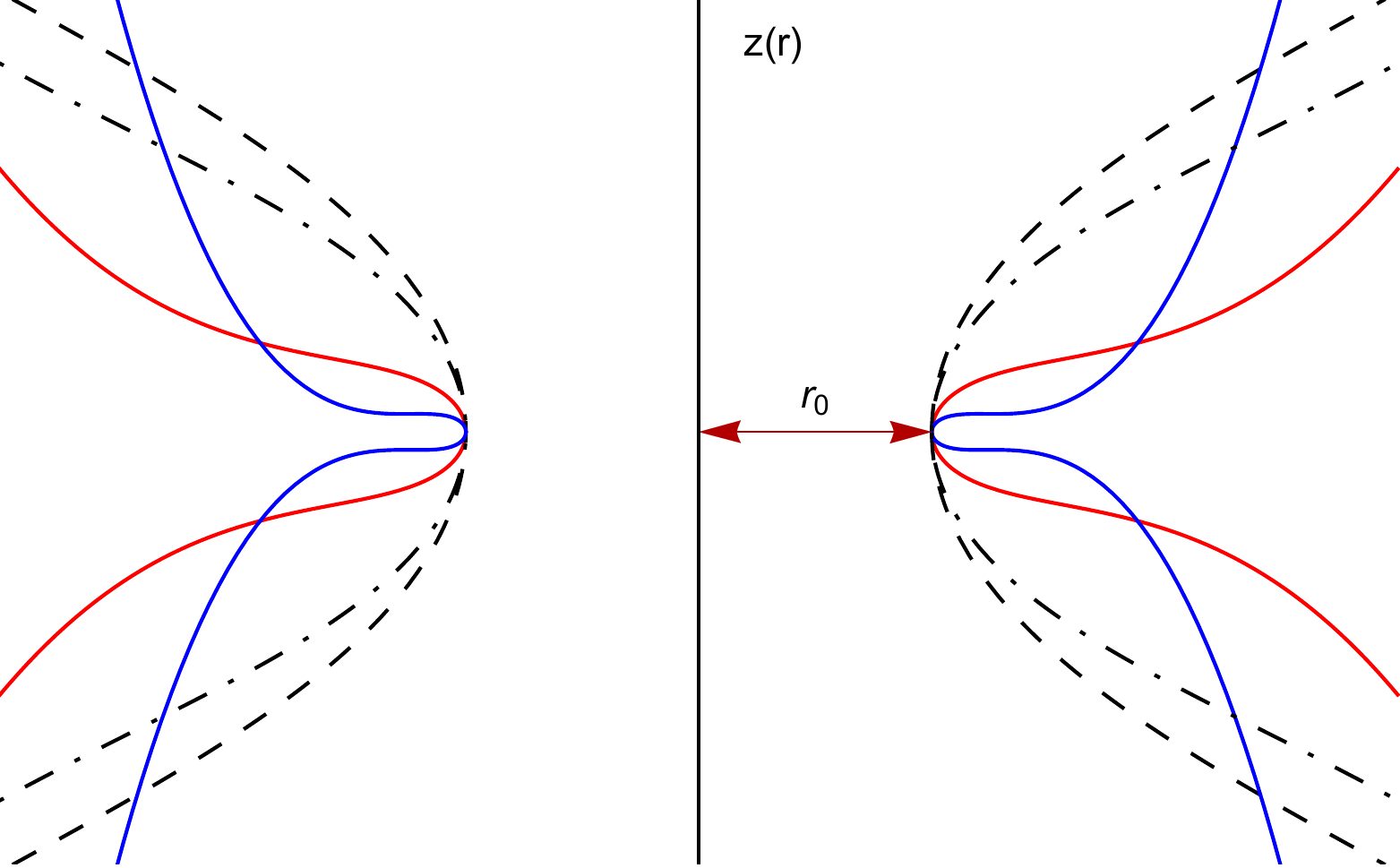}
		\caption{Embedding function against radial coordinate for zero (left panel) and nonzero (right panel) redshift solutions. The model parameters have been set according to Figs.~(\ref{fig3}) and~(\ref{fig5}).}\label{fig8}
	\end{center}
\end{figure}

The radial coordinate $r$ exhibits non-monotonic behavior, where in, it decreases from $+\infty$ to a minimum value $r=r_0$ i.e., the location of the wormhole throat where $b(r_0) = r_0$. It then increases from $r_0$ to $+\infty$. Although the radial metric component i.e., $g_{rr}$ diverges in the limit $r\rightarrow r_0$, the proper radial distance~\cite{mt,FLoboBook}
\be\label{properl}
l(r)=\pm\int_{r_0}^{r}\f{dy}{\sqrt{1-\f{b(y)}{y}}},
\ee
is required to be finite everywhere. This quantity decreases from $\ell=+\infty$ in the upper Universe to $\ell=0$ at the throat, and then from zero to $\ell=-\infty$ in the lower Universe. We can compute the above integral for the zero redshift solution, Eq.~(\ref{solb}) and the result can be expressed in terms of hypergeometric functions as
\bea\label{ellhyper}
&&\ell(r)=\f{\pm1}{\sqrt{2}}\Bigg\{\frac{\sqrt{\pi}\Gamma\left[\frac{1}{m+4}\right]}{\sqrt{\f{(m+4)^2(w+1)}{r_0^2(3w-1)(2\omega+3)}}\Gamma \left[\f{1}{2}+\f{1}{m+4}\right]}+\f{r(3w-1)(2\omega+3)}{w+1}\left[\f{(1+w)(r_0^{m+4}-r^{m+4})}{(2\omega+3)(3w-1)r_0^{m+4}}\right]^{\f{1}{2}}\times\nn&&{_2}F_1\left[1,\f{1}{2}+\f{1}{m+4},1+\f{1}{m+4},\left(\f{r}{r_0}\right)^{m+4}\right]\Bigg\},
\eea
where the integration constant has been set in such a way that $\ell(r_0)=0$ and use has been made of the expressions given in Eqs.~(\ref{bdsfr0}) and (\ref{subs}). For the case of nonzero tidal force solutions, we may substitute Eq.~(\ref{shapefin}) into Eq.~(\ref{properl}) to find the proper distance, however, the resulted integral cannot be solved in terms of standard mathematical functions. We therefore proceed to consider the Taylor expansion of the integrand around the throat which gives
\bea\label{lpropnonex}
\ell(r)&=&\pm\int_{r_0}^{r}dy\left(1-\f{b(y)}{y}\right)^{-\f{1}{2}}\!\Bigg|_{r=r_0}\approx\f{2\sqrt{r-r_0}}{\left[\f{wr_0-\gamma(1+w)}{wr_0^2}\right]^{\f{1}{2}}}+\f{\left[\f{wr_0-\gamma(1+w)}{wr_0^2}\right]^{\f{1}{2}}\delta}{6\left[wr_0-\gamma(1+w)\right]^2}\left(r-r_0\right)^{\f{3}{2}}\nn&+&\f{\left[\f{wr_0-\gamma(1+w)}{wr_0^2}\right]^{\f{3}{2}}\epsilon_1}{240\left[w\gamma+\gamma-wr_0\right]^4}\left(r-r_0\right)^{\f{5}{2}}+{\mathcal O}\left(r-r_0\right)^{\f{7}{2}},
\eea
where 
\bea\label{delel}
\delta&=&2w^2r_0^2-4wr_0\gamma(1+w)+\gamma^2(1+w)^2,\nn
\epsilon_1&=&-12r_0^4 w^4+48 \gamma r_0^3(w+1)w^3-36\gamma^2r_0^2(w+1)^2w^2+8\gamma^3r_0(w+1)^3w+\gamma^4(w+1)^4.
\eea
The behavior of the proper distance for zero tidal force solutions and those of nonzero tidal force has been plotted in Fig.~(\ref{fig7}). One may also seek for visualizing the obtained wormhole solutions through embedding diagrams. To this aim, we may firstly consider a three-dimensional space at a fixed moment of time, i.e., setting $t=const.$ within Eq.~(\ref{metric}). As the geometry is spherically symmetric, then without loss of generality, we can restrict ourselves to the equatorial slice at which $\theta=\pi/2$. Therefore, the two-dimensional line element is found as
\be\label{mteric2d}
ds^2=\left[1-\f{b(r)}{r}\right]^{-1}dr^2+r^2d\phi^2.
\ee
Now, in order to get a picture of the above slice we can embed it into a three-dimensional Euclidean space for which, the line element in cylindrical coordinates $(r,\phi,z)$ is written as
\be\label{lineelEu}
ds^2=dz^2+dr^2+r^2d\phi^2.
\ee
In the three-dimensional Euclidean space the embedded surface will be axially symmetric and thus can be represented by the single function $z=z(r)$. The line element on this surface then reads 
\be\label{linellsurf}
ds^2=\left[1-\left(\f{dz(r)}{dr}\right)^2\right]dr^2+r^2d\phi^2,
\ee 
where the function $z(r)$ describes the embedded surface. If we consider the coordinates $(r,\phi)$ of the embedding space same as those of the wormhole's spacetime and match the two line elements, we arrive at the following differential equation (and the corresponding integral) for the embedding function 
\be\label{zfunct}
\f{dz}{dr}=\pm\left[\f{r}{b(r)}-1\right]^{-\f{1}{2}},~~~~~~~~~~~~~~z(r)=\pm\int_{r_0}^{r}\f{dx}{\sqrt{\f{x}{b(x)}-1}}.
\ee
From the above expression and radial proper distance Eq.~(\ref{properl}) we may recognize that
\be\label{zrthr}
\f{dz}{d\ell}=\pm\sqrt{\f{b}{r}}~~\Rightarrow~~~\f{dz}{d\ell}\Bigg|_{r\rightarrow r_0}\!\!\!\!\!\!\!\!\!=~\pm1~~\Rightarrow~~ z(r_0)=\pm\ell(r_0)=0,~~~~~~{\rm and}~~~~~\f{dz}{dr}\Bigg|_{r\rightarrow r_0}\!\!\!\!\!\!\!\!\!=~\pm\infty.
\ee
Now, substituting the zero redshift solution into Eq.~(\ref{zfunct}) we obtain the embedding function as the following form
\bea
z(r)&=&\f{\pm1}{\sqrt{2}}\Bigg[\f{r\left[1-\left(\f{r}{r_0}\right)^{m+4}\right]^{\f{1}{2}}F_1\left[\f{1}{m+4},\frac{1}{2},-\frac{1}{2},\f{m+5}{m+4},\left(\f{r}{r_0}\right)^{m+4},\f{2(w+1)}{2\omega-w(6\omega+7)+5}\left(\f{r}{r_0}\right)^{m+4}\right]}{\sqrt{\f{2(w+1) r^{m+4}}{[w(6\omega+7)-2\omega-5]r_0^{m+4}}+1} \sqrt{\f{(w+1)\left(r_0^{m+4}-r^{m+4}\right)}{2 (w+1) r^{m+4}+r_0^{m+4} (w (6 \omega +7)-2 \omega -5)}}}\nn&-&\f{\sqrt{\pi } r_0 \sqrt{-\f{m+4}{r_0}}\,\Gamma\!\left[\frac{m+5}{m+4}\right]{_2}\tilde{F}_1\left[-\f{1}{2},\frac{1}{m+4},\frac{m+6}{2(m+4)},\f{2 w+2}{-6 \omega  w-7 w+2 \omega +5}\right]}{\sqrt{\f{2 (w+1)}{w (6 \omega +7)-2 \omega -5}+1} \sqrt{\f{(m+4) (w+1)}{r_0(1-3w) (2 \omega +3)}}}\Bigg],
\eea
where $F_1$ is the Appell hypergeometric function of two variables and $_2\tilde{F}_1$ is the regularized hypergeometric function~\cite{AppellHyper}. For the case of nonzero redshift solutions the integral Eq.~(\ref{zfunct}) cannot be solved in a straightforward way. We therefore try near throat approximation which gives
\bea\label{zfunnonex}
z(r)\!\!\!\!&=&\!\!\!\!\pm\int_{r_0}^{r}dx\left(\f{x}{b(x)}-1\right)^{-\f{1}{2}}\Bigg|_{r=r_0}\!\!\!\!\!\!\!\!\approx\f{2}{\left[\f{wr_0-\gamma(1+w)}{wr_0^2}\right]^{\f{1}{2}}}(r-r_0)^{\f{1}{2}}-\f{\gamma^2(1+w)^2\left[\f{wr_0-\gamma(1+w)}{wr_0^2}\right]^{\f{1}{2}}}{6(\gamma w+\gamma-wr_0)^2}(r-r_0)^{\f{3}{2}}\nn&+&\f{\gamma^2(1+w)^2\left[\f{wr_0-\gamma(1+w)}{wr_0^2}\right]^{\f{1}{2}}\epsilon_2}{240(w\gamma+\gamma-wr_0)^4}(r-r_0)^{\f{5}{2}}+{\mathcal O}(r-r_0)^{\f{7}{2}},
\eea
where
\be\label{ep2}
\epsilon_2=24r_0^2 w^2-16 \gamma r_0(w+1) w+\gamma ^2 (w+1)^2.
\ee
Figure~(\ref{fig8}) shows the embedding diagram for the wormhole solutions. In the left panel we have sketched the solutions with $\Phi=0$ and those related to $\Phi\neq0$ are shown in the right panel. The left side is the mirror image of the embedding diagram on the right. We therefore observe that the shape of wormhole in the vicinity of the throat can be affected by a zero or nonzero redshift function.
\subsection{Spacetime singularities}
One of the most prominent and longstanding issues in GR and in metric theories of gravity, such as BD theory, is the unavoidable occurrence
of spacetime singularities in physically relevant solutions of GR field equation or the BD theory, see e.g.,~\cite{singgrbd,singgrbd1}. A singularity is a spacetime event at which densities as well as curvature invariants grow unboundedly and diverge. The occurrence of such an event is accompanied by important implications for the issues such as physical determinism, predictability, causality and validity of the standard laws of physics~\cite{singuissues}. Spacetime singularities are commonly thought to be problematic since the classical framework of GR fails to be predictive and thus loses its application when basic quantities become infinite. Hence, formation of a spacetime singularity signals that the theory has been employed beyond its domain of validity~\cite{GRSINDOM,GRSINDOM1}. The fundamental singularities of a spacetime can be classified into three major groups: quasi-regular, non-scalar curvature and scalar curvature singularities. An example of the first one is the conical singularity in the spacetime of the cosmic string~\cite{GRSINDOM1}. The second category is referred to as singularities whose curvature scalars do not badly behave~\cite{Ellis1977}. These singularities are also called matter singularities if the problem arises with some component of the Ricci tensor~\cite{GRSINDOM1}. Scalar curvature singularities are the end point of at least one causal curve on which curvature invariants such as Kretschmann scalar diverge. As examples of these type of singularities one may quote the inner Schwarzschild solution and the initial singularity of the Universe, see e.g.,~\cite{Ellis1977},\cite{sing1979} for more details. It is generally accepted that the geodesic incompleteness and also divergence of the curvature invariants strongly indicate the presence of a spacetime singularity~\cite{singgrbd,Ellis1977,KRSING,Racz2023}. Regarding these considerations, it can be argued that the problem of spacetime singularities is also important in wormhole physics. If the singularity is present within the spacetime of one of the Universes that are connected through the wormhole, then the pathological behavior of extreme regions (e.g., breakdown of physical laws) at the neighborhood of the singularity can be transferred from one Universe to the other or, in case in which the throat is singular, such effects may be transferred to both Universes which are connected by the wormhole throat. This can happen if we have a timelike singularity e.g., at the throat to which a test particle can get close arbitrarily and then escape the singular region. If so, the test particle may carry the fundamental problems mentioned above with it to both Universes and disrupt the known laws of physics elsewhere~\cite{singuissues}. Consequently, for a physically reasonable wormhole configuration, we require that the spacetime singularities are absent at the throat and throughout the spacetime. To this aim we proceed to compute the Kretschmann invariant which for the line element (\ref{metric}) takes the following form
\bea\label{Kretschamnn}
\!\!\!\!\!\!\!\!\!{\mathcal K}\!\!\!\!&=&\!\!\!\!R^{\mu\nu\alpha\beta}R_{\mu\nu\alpha\beta}=\!\f{1}{r^6}\Bigg[4 r^3 (r-b) \left(b-r b'\right) \Phi'^3+8 r^3 (r-b) \Phi'\Phi''\left[\f{b}{2}-\f{1}{2} r b'+r (r-b) \Phi'\right]\nn\!\!\!\!\!\!\!\!\!\!&+&\!\!\!\!r^2\Big(r^2 b'^2-2 rb b'-16 r b+9 b^2+8 r^2\Big) \Phi'^2+2 r^2b'^2-4 r bb'+4 r^4 (r-b)^2\left[\Phi'^4+\Phi''^2\right]+6b^2\Bigg].
\eea
We firstly note that matter distribution behaves regularly at the throat as, all of the EMT components are finite in the limit of approach to the throat, i.e., a singularity free EMT. The Kretschmann scalar for zero and nonzero tidal force solutions is obtained as
\bea\label{KRzNZ}
{\mathcal K}_{\rm zero}&=&k_1\left[\f{r}{r_0}\right]^{2m+4}+\f{k_2}{r^4}+k_3\left[\f{r}{r_0}\right]^m,\nn
{\mathcal K}_{\rm nonzero}&=&\f{k_4}{r^8}+\f{k_5}{r^9}{\rm exp}\left[\f{(r-r_0)(1+w)\gamma}{wrr_0}\right]+\f{k_6}{r^{10}}{\rm exp}\left[\f{2(r-r_0)(1+w)\gamma}{wrr_0}\right],
\eea
where
\bea\label{kcoefs}
k_1&=&\f{8(1+w)^2[m(8+m)+18]}{r_0^4(2\omega+3)^2(1-3w)^2},~~~~~~k_2=\f{4 [2\omega+5-w (6 \omega +7)]^2}{(1-3 w)^2 (2 \omega +3)^2},\nn
k_3&=&\f{16 (w+1) [w (6 \omega +7)-2 \omega -5]}{r_0^4(1-3 w)^2 (2 \omega +3)^2},~~~~~~~~k_4=4\gamma^2\left(6r^2+4r\gamma+\gamma^2\right),\nn
k_5&=&\f{4 \gamma ^2 r_0}{w}\left[\gamma r(2-7w)+\gamma^2(1-w)-14 r^2 w\right],\nn
k_6&=&\f{r_0^2}{w^2}\left[6r^4w^2-4\gamma  r^3w(w+1)+\gamma^2r^2[w (35 w+4)+2]+10\gamma^3r(w-1)w+\gamma^4(w-1)^2\right].
\eea
The above quantities at the throat will take the form 
\bea\label{KRet0}
{\mathcal K}_{\rm zero}\Big|_{r=r_0}&=&\f{4}{r_0^4}\left[\frac{2 (m+4)^2 (w+1)^2}{(1-3 w)^2 (2 \omega +3)^2}+1\right],\nn
{\mathcal K}_{\rm nonzero}\Big|_{r=r_0}&=&\frac{\gamma ^4 (w+1)^2}{r_0^8 w^2}-\frac{2 \gamma ^3 (w+1)}{r_0^7 w}+\frac{\gamma^2[w (3 w+4)+2]}{r_0^6 w^2}-\frac{4 \gamma  (w+1)}{r_0^5 w}+\frac{6}{r_0^4}.
\eea
From the above expressions we find out that if $w\neq0$, the Kretschmann invariant is finite at the throat for the solutions with nonvanishing redshift function. Also, for $w\neq-1/3~\land~\omega\neq-3/2$, this quantity behaves regularly at wormhole throat for the class of zero redshift solutions. As a result, there is no curvature singularity at the throat. We further note that these quantities are well-behaved for $r>r_0$ and tend to zero asymptotically. Hence, a curvature singularity is absent throughout the wormhole spacetime. 
\par
\section{Concluding Remarks}\label{concluding}
The study of wormhole configurations in BD theory has attracted many interests in recent decade since in this theory, the scalar field can play the role of the exotic matter at the wormhole throat and ensures it’s traversability~\cite{bd},\cite{bd1}. In the present work we tried to find exact wormhole solutions in BD theory in non-vacuum spacetimes. As the standard conservation law holds in BD theory, i.e., Eq.~(\ref{conseq}), we utilized this equation along with a linear EoS between radial pressure and energy density in order to arrive at the system of differential equations Eqs.~(\ref{bdevoltov}) and (\ref{bdrhowpr}). For this system, two classes of solutions were generally achieved. In the first one we dealt with zero tidal force solutions presented in Sec.~(\ref{redzero}), for which, we examined the conditions on wormhole geometry as well as NEC and WEC. These conditions put restrictions on model parameters which for example, depending on the value of energy density at wormhole throat and the throat radius, one finds the allowed regions of the space parameter as shown in the left panel of Fig.~(\ref{fig1}). For the second class, we performed the same strategy for nonzero tidal force solutions, see Sec.~(\ref{rednzero}), and obtained the space parameter Fig.~(\ref{fig4}) accordingly. In the first part of Sec.~(\ref{sofuis}), we obtained Schwarzschild-like as well as MT wormhole solutions for particular values of model parameters. We further obtained constraints on model parameters subject to the bounds on BD coupling parameter, previously reported in the literature. In the second part, we considered the proper radial distance along with the embedding function and found that the shape of the wormhole in the neighborhood of the throat depends crucially on whether the redshift function is zero or not. Finally, in the third part, we investigated possible occurrence of curvature singularities in the wormhole spacetime. This is an important issue in gravitation and cosmology since, in the spacetimes admitting naked singularities we may not be able to determine regular initial data in the presence of a naked singularity and consequently the initial value problem fails. This makes the theory loses its predictability and be unphysical, see e.g.,~\cite{singuissues,GRSINDOM,GRSINDOM1},\cite{valerio2016,SPR1393} and references therein. Regarding these considerations we found that, in non-vacuum BD spacetimes, static spherically symmetric regular wormhole configurations (without spacetime singularity at the throat or beyond it) can exist that comply with the NEC as well as WEC. {Our findings could shed some light on the idea that the BD theory may naturally accommodate wormhole spacetimes supported by ordinary matter sources, as the non-minimal coupling of the BD scalar field can effectively mimic exotic matter~\cite{bd,bd1,brans1962,bd2,bd4,bd3,bd3rev,bd5,bd6,bd7}. This may contribute to the ongoing debate about whether the additional degrees of freedom in modified gravity theories can play the role of exotic matter, providing thus a suitable setting for wormhole structures that respect WEC and NEC, the scenario which is absent in GR~\cite{FLoboBook,lobocgqreview}. Paging through the literature, one can also find a variety of research works that deal with the role of exotic matter in creating traversable wormholes and discuss its implications for different modified gravity theories, see e.g.,~\cite{lobocgqreview},\cite{fr1},\cite{otherworms,newast2024}, \cite{Liu2024}. The strategy used here for solving the coupled field equations could inspire similar constructions in more general scalar-tensor theories such as Horndeski gravity~\cite{Hornworm} or Einstein-scalar-Gauss-Bonnet theory~\cite{gmfl}, where wormholes or other exotic solutions remain underexplored. As the final remark, it is worth mentioning that, exotic spacetimes in BD theory may serve as toy models for quantum gravity, as the scalar field could encode quantum corrections at semi-classical levels~\cite{semibdquant}, see also \cite{bd1loop} where the authors discussed that the Brans-Dicke model appears as the 1-loop approximation to the string theory. Hence, BD wormholes may provide a compelling bridge between classical modified gravity and the quantum realm, while inviting future investigations into deeper theoretical connections of these intriguing spacetime geometries~\cite{FLoboBook,khu},\cite{quantumworm}.}
\par
\vspace{0.5cm}
{\bf Data Availability Statement}: The data that support the findings of this study are available from the corresponding author, [AHZ], upon reasonable request.
\section*{Acknowledgement} 
We would like to appreciate the anonymous referees for providing useful and constructive comments that helped us to improve the original version of our manuscript.


\begin{thebibliography}{99}
\bibitem{FLoboBook} F. S. N. Lobo (Editor), \lq{}\lq{}{\it Wormholes, Warp Drives and Energy Conditions,}\rq{}\rq{} Springer (2017).
\bibitem{khu} M. Visser, \lq{}\lq{}{\it Lorentzian Wormholes: From Einstein to Hawking}\rq{}\rq{} (AIP, Woodbury, USA, 1995).
\bibitem{Flamm} L. Flamm, Phys. Z. {\bf 17} 448 (1916);\\G. W. Gibbons, Editorial note to: Ludwig Flamm, Republication of: contributions to Einsteins theory of gravitation. Gen. Relativ. Gravit. {\bf 47}, 72 (2015).
\bibitem{ERose} A. Einstein and N. Rosen, Phys. Rev. {\bf 48}, 73 (1935);\\D. R. Brill and R. W. Lindquist, Phys. Rev. {\bf 131}, 471 (1963).
\bibitem{misner-wheeler} C. W. Misner and J. A. Wheeler, Ann. Phys. {\bf 2}, 525 (1957); \\C. W. Misner, Phys. Rev. {\bf 118}, 1110 (1960).
\bibitem{Wheelerworm} J. A. Wheeler, Ann. Phys. {\bf 2}, 604 (1957);\\ J. A. Wheeler, \lq{}\lq{}{\it Geometrodynamics,}\rq{}\rq{} (Academic, New York, 1962).
\bibitem{GrBrRN} J. C. Graves \& D. R. Brill, Phys. Rev. {\bf 120}, 1507 (1960).
\bibitem{Homers} H. G. Ellis, J. Math. Phys. {\bf 14}, 104 (1973); Gen. Relativ. Gravit. {\bf 10}, 105 (1979).
\bibitem{Bronnsols} K. A. Bronnikov, Acta Phys. Polon. B {\bf 4}, 251 (1973).
\bibitem{hisworm} F. S. N. Lobo, Int. J. Mod. Phys. D, {\bf 25}, 1630017 (2016).
\bibitem{mt} M. S. Morris and K. S. Thorne, Am. J. Phys. {\bf 56}, 395
(1988).
\bibitem{mt1} M. S. Morris, K. S. Thorne and U. Yurtsever, Phys. Rev. Lett. {\bf 61}, 1446 (1988).
\bibitem{VissHoch1997} M. Visser \& D. Hochberg, arXiv:gr-qc/9710001.
\bibitem{khu1} D. Hochberg and M. Visser, Phys. Rev. D {\bf 56}, 4745 (1997).
\bibitem{Drumm2004} P. D. Drummond \& Z. Ficek, (eds.), \lq{}\lq{}{\it Quantum Squeezing,}\rq{}\rq{} Springer-Verlag, Berlin (2004).
\bibitem{Gsqueeze} D. Hochberg \& T. W. Kephart, Phys. Lett. B {\bf 268} 377 (1991).
\bibitem{qftcasm} B. S. DeWitt, Phys. Rep. {\bf 19} 295 (1975);\\B. S. DeWitt, \lq{}\lq{}{\it Quantum gravity: the new synthesis,}\rq{}\rq{} General Relativity: An Einstein Centenary Survey, edited by S. W. Hawking and W. Israel, Cambridge Univ. Press, Cambridge (1979);\\N. D. Birrell \& P. C. W. Davies, \lq{}\lq{}{Quantum fields in curved space,}\rq{}\rq{}, Cambridge University Press, Cambridge (1984);\\K. A. Milton, \lq{}\lq{}{\it The Casimir Effect: Physical Manifestations of Zero-point Energy,}\rq{}\rq{} Singapore: World Scientific (2001).
\bibitem{edavis2012} E. W. Davis, An Assessment of Faster-Than-Light Spacetimes: Make or Break Issues, \url{https://doi.org/10.2514/6.2006-4908};\\
E. W. Davis \& H. E. Puthoff, Experimental Concepts for Generating Negative Energy in the Laboratory, AIP Conf. Proc. {\bf 813}, 1362 (2006).
\bibitem{exo0} S. W. Hawking, Phys. Rev D {\bf 46}, 603 (1992).
\bibitem{exo1} E. Poisson and M. Visser, Phys. Rev. D {\bf 52}, 7318 (1995).
\bibitem{exo2} M. Chianese, E. Di Grezia, M. Manfredonia and G. Miele, Eur. Phys. J. Plus {\bf 132}, 164 (2017).
\bibitem{phantworm} F. S. N. Lobo, Phys. Rev. D {\bf 71}, 124022 (2005); \\P. K. F. Kuhfittig, Class. Quant. Grav. {\bf 23}, 5853 (2006); \\F. S. N. Lobo, F. Parsaei and N. Riazi, Phys. Rev. D {\bf 87}, 084030 (2013);\\ R. Garattini, Eur. Phys. J. C {\bf 79}, 951 (2019);\\R. Garattini and A. G. Tzikas, JCAP 12, 019 (2024).
\bibitem{intdarksec} V. Folomeev and V. Dzhunushaliev, Phys. Rev. D {\bf 89}, 064002 (2014);\\{S. Chaudhary, S. K. Maurya, J. Kumar, S. Kiroriwal, Pramana J. Phys. {\bf 98}, 139 (2024).}
\bibitem{cutpaste} M. Visser, Phys. Rev. D {\bf 39}, 3182(R) (1989); Nucl. Phys. B {\bf 328}, 203 (1989);\\F. S. Lobo, P. Crawford, Class. Quantum Gravity {\bf 21}, 391 (2003).
\bibitem{thinshells} M. Bhatti, M. Yousaf, Z. Yousaf, New Astron. {\bf 106}, 102132 (2024);\\J. P. Lemos, F. S. Lobo, Phys. Rev. D {\bf 78}, 044030 (2008);\\F. Javed, S. Mumtaz, G. Mustafa, I. Hussain, W.-M. Liu, Eur. Phys. J. C {\bf 82}, 825 (2022);\\G. Mustafa, S. Maurya, S. Ray, F. Javed, Ann. Phys. {\bf 460}, 169551 (2024).
\bibitem{minexot} S. Kar, N. Dadhich, M. Visser, Pramana {\bf 63}, 859 (2004);\\E. F. Eiroa and C. Simeone, Phys. Rev. D {\bf 71}, 127501 (2005);\\O. B. Zaslavskii Phys. Rev. D {\bf 76}, 044017 (2007);\\M. Bouhmadi-Lopez, F. S. N. Lobo, P. Martin-Moruno, JCAP {\bf 1411},  007 (2014);\\ F. Parsaei \& S. Rastgoo, Eur. Phys. J. C {\bf 80}, 366 (2020);\\F. Parsaei \& S. Rastgoo, Phys. Rev. D {\bf 99}, 104037 (2019).
\bibitem{lobocgqreview} F. S. N. Lobo, Classical and Quantum Gravity Research, 1-78, (2008), Nova Sci. Pub. ISBN 978-1-60456-366-5, arXiv:0710.4474 [gr-qc].
\bibitem{modsgrex} F. S. N. Lobo, Class. Quantum Grav. {\bf 25}, 175006 (2008);\\C. G. Boehmer, T. Harko, F. S. N. Lobo, Phys. Rev. D {\bf 85}, 044033 (2012);\\T. Harko, F. S. N. Lobo, M. K. Mak, and S. V. Sushkov, Phys. Rev. D {\bf 87}, 067504 (2013);\\F. Duplessis and D. A. Easson, Phys. Rev. D {\bf 92}, 043516 (2015);\\M. K. Zangeneh, F. S. N. Lobo, M. H. Dehghani, Phys. Rev. D {\bf 92}, 124049 (2015);\\G. U. Varieschi and K. L. Ault, Int. J. Mod. Phys. D {\bf 25} 1650064 (2016);\\M. Hohmann, C. Pfeifer, M. Raidal, H. Veerm\"{a}e, JCAP 1810 (2018) no.10, 003;\\M. Zubair, F. Kousar, S. Bahamonde, Eur. Phys. J. Plus {\bf 133} 523 (2018);\\G. C. Samanta, N. Godani, Eur. Phys. J. C {\bf 79}, 623 (2019);\\A. \"{O}vg\"{u}n, K. Jusufi, I. Sakalli, Phys. Rev. D {\bf 99}, 024042 (2019).
\bibitem{higherdimw} A. Chodos and S. Detweiler, Gen. Rel. Grav. {\bf 14}, 879 (1982);\\G. Cl\'{e}ment, Gen. Rel. Grav. {\bf 16}, 131 (1984);\\A. De Benedictis and A. Das, Nucl. Phys. B {\bf 653}, 279 (2003).
\bibitem{KKGworm} V. Dzhunushaliev, H.-J. Schmidt, Phys. Rev. D {\bf 62}, 044035 (2000);\\J. P. de Leon, J. Cosmol. Astropart. Phys. 2009 (11), 013 (2009);\\V. Dzhunushaliev and V. Folomeev, Mod. Phys. Lett. A {\bf 29}, 1450025 (2014);\\K. Sarkar, G. Biswas, B. Modak, Gen. Relativ. Gravit. {\bf 50}, 157 (2018).
\bibitem{gmfl} P. Kanti, B. Kleihaus, J. Kunz, Phys. Rev. Lett. {\bf 107}, 271101 (2011);\\ S. H. Mazharimousavi, M. Halilsoy, and Z. Amirabi, Phys.
Rev. D {\bf 81}, 104002 (2010);\\ P. Kanti, B. Kleihaus and J. Kunz, Phys. Rev. D {\bf 85}, 044007 (2012);\\ P. Canate, N. Breton, Phys. Rev. D {\bf 100}, 064067 (2019);\\
G. Antoniou, A. Bakopoulos, P. Kanti, B. Kleihaus and J. Kunz, Phys. Rev. D {\bf 101}, 024033 (2020);\\ R. Ibadov, B. Kleihaus, J. Kunz, S. Murodov, Phys. Rev. D {\bf 102}, 064010 (2020).
\bibitem{braneworm} L. A. Anchordoqui and S. E. Perez Bergliaffa, Phys. Rev. D {\bf 62}, 076502 (2000);\\C. Barcel\'{o} and M. Visser, Nucl. Phys. B {\bf 584}, 415 (2000);\\K. A. Bronnikov and S.-W. Kim, Phys. Rev. D {\bf 67}, 064027 (2003).
\bibitem{LOVEWORM} G. Dotti, J. Oliva, R. Troncoso, Phys. Rev. D {\bf 75}, 024002 (2007);\\
H. Maeda, M. Nozawa, Phys. Rev. D {\bf 78}, 024005 (2008);\\M. H. Dehghani and Z. Dayyani, Phys. Rev. D {\bf 79}, 064010 (2009);\\M. R. Mehdizadeh and F. S. N. Lobo,  Phys. Rev. D {\bf 93}, 124014 (2016);\\G. Giribet, E. R. de Celis, and C. Simeone, Phys. Rev. D {\bf 100}, 044011 (2019).
\bibitem{fr} N. Furey and A. DeBenedictis, Class. Quantum Grav. {\bf 22}, 313 (2005);\\ F. S. N. Lobo and M. A. Oliveira, Phys. Rev. D {\bf 80}, 104012 (2009); \\A. De Benedictis, D. Horvat, Gen. Relat. Gravit. {\bf 44}, 2711 (2012);\\S. Bhattacharya, S., Chakraborty, Eur. Phys. J. C {\bf 77}, 558 (2017);\\M. Sharif and I. Nawazish, Annals of Physics, {\bf 389}, 283 (2018).
\bibitem{fr1} O. Sokoliuk, S. Mandal, P. K. Sahoo, A. Baransky, Eur. Phys. J. C {\bf 82}, 280 (2022);\\R. Radhakrishnan, P. Brown, J. Matulevich, E. Davis, D. Mirfendereski, G. Cleaver, Symmetry, {\bf 16}, 1007 (2024).
\bibitem{massiveworm} {S. I. Vacaru, Eur. Phys. J. C {\bf 74}, 2781 (2014);\\T. Tangphati, A. Chatrabhuti, D. Samart, P. Channuie, Eur. Phys. J. C {\bf 80} 722 (2020);\\A. Dutta, D. Roy, N. J. Pullisseri, S. Chakraborty, Eur. Phys. J. C {\bf 83}, 500 (2023);\\ J. Kumar, S. K. Maurya, S. Kiroriwal, Eur. Phys. J. C {\bf 84}, 1305 (2024);\\S. Kiroriwal, S. K. Maurya, J. Kumar, A. Errehymy, G. Mustafa, K. Myrzakulov, Chinese J. Phys.	{\bf 95}, 404 (2025);\\ J. Kumar, S. K. Maurya, S. Kiroriwal, A. Errehymy, O. Donmez, K. Myrzakulov, J. High Energy Astrophys. {\bf 45}, 32 (2025).}
\bibitem{ecworm} K. A. Bronnikov, A. M. Galiakhmetov, Grav. Cosmol, {\bf 21}, 283 (2015); Phys. Rev. D {\bf 94}, 124006 (2016);\\ M. R. Mehdizadeh, A. H. Ziaie, Phys. Rev. D {\bf 95}, 064049 (2017); Phys. Rev. D {\bf 99}, 064033 (2019).
\bibitem{Garcia-Lobo} N. M. Garcia and F. S. N. Lobo, Phys. Rev. D {\bf 82}, 104018 (2010);\\ M. Zubair, S. Waheed and Y. Ahmad, Eur. Phys. J. C {\bf 76}, 444 (2016);\\ N. Godani and G. C. Samanta, Chinese J. Phys. {\bf 62}, 161 (2019);\\R. Solanki, Z. Hassan, P. K. Sahoo, Chin. J. Phys. {\bf 85} 74 (2023);\\ {T. Naseer, M. Sharif, A. Fatima, S. Manzoor, Chinese J. Phys. {\bf 86}, 350 (2023)};\\N. Loewer, M. Tayde, P. K. Sahoo, Eur. Phys. J. C {\bf 84}, 1196 (2024);\\ {T. Naseer, M. Sharif, M. Faiza, B. Dayanandan, Eur. Phys. J. C {\bf 84}, 1187 (2024)};\\{T. Naseer, M. Sharif, M. Faiza, Chinese J. Phys. {\bf 94}, 204 (2025);\\ T. Naseer, M. Sharif, M. Faiza, W. Albalawi, A.-H. Abdel-Aty, Phys. Dark Univ. {\bf 48}, 101890 (2025).}
\bibitem{rastallworm} H. Moradpour, N. Sadeghnezhad and S. H. Hendi, Can. J. Phys. {\bf 95}, 1257 (2017);\\S. Halder, S. Bhattacharya, and S. Chakraborty, Mod. Phys. Lett. A {\bf 34}, 1950095 (2019);\\I. P. Lobo, M. G. Richarte, J. P. Morais Graça, H. Moradpour, Eur. Phys. J. Plus {\bf 135} 550 (2020);\\G. Mustafa, M. R. Shahzad, G. Abbas, and T. Xia, Mod. Phys. Lett. A {\bf 35}, 2050035 (2020).
\bibitem{otherworms} R. Shaikh, Phys. Rev. D {\bf 92}, 024015 (2015);\\ F. Rahaman, N. Paul, A. Banerjee, S. S. De, S. Ray and A. A. Usmani, Eur. Phys. J. C {\bf 76}, 246 (2016);\\ P. H. R. S. Moraes, P. K. Sahoo, Phys. Rev. D {\bf 96}, 044038 (2017);\\ M. G. Richarte, I. G. Salako, J. P. Morais Graca, H. Moradpour, and A. ovgun, Phys. Rev. D {\bf 96}, 084022 (2017);\\K. Jusufi, N. Sarkar, F. Rahaman, A. Banerjee and S. Hansraj, Eur. Phys. J. C {\bf 78} 349 (2018);\\S. Kiroriwal, J. Kumar, S. K. Maurya, S. Chaudhary, Eur. Phys. J. C {\bf 84}, 414 (2024);\\
P. H. F. Oliveira, G. Alencar, I. C. Jardim, R. R. Landim, Mod. Phys. Lett. A. {\bf 37}, 2250090 (2022);\\ F. Parsaei, S. Rastgoo, P. K. Sahoo, Eur. Phys. J. Plus {\bf 137}, 1083 (2022);\\A. S. Agrawal, S. Zerbini, B. Mishra, Phys. Dark Univ. {\bf 46}, 101637 (2024);\\{S. Kiroriwal, J. Kumar, S. K. Maurya, S. Chaudhary, A. Aziz, Fortschritte der Physik, {\bf 72}, 2300197 (2024);\\ S. Chaudhary, S. K. Maurya, J. Kumar, S. Ray, Astroparticle Physics, {\bf 162}, 103002 (2024);\\ S. Chaudhary, J. Kumar, S. K. Maurya, S. Kiroriwal, A. Aziz, Commun. Theor. Phys. {\bf 76} 055403 (2024);\\ J. Kumar, S. K. Maurya, S. Kiroriwal, A. Errehymy, K. Myrzakulov, Z. Umbetova, Phys. Dark Univ. {\bf 46}, 101636 (2024).}
\bibitem{newast2024} {J. Kumar, S. K. Maurya, S. Kiroriwal, S. Chaudhary, New Astro. Rev. {\bf 98}, 101695 (2024).}
\bibitem{BDTH} C. Brans and R. H. Dicke, Phys. Rev. {\bf 124}, 925 (1961).
\bibitem{bd} A. G. Agnese and M. La Camera, Phys. Rev. D {\bf 51}, 2011 (1995).
\bibitem{bd1} K. K. Nandi, A. Islam, and J. Evans, Phys. Rev. D {\bf 55}, 2497 (1997).
\bibitem{brans1962} C. H. Brans, Phys. Rev. {\bf 125}, 2194 1962.
\bibitem{bd2} F. He and S.-W. Kim, Phys. Rev. D {\bf 65}, 084022 (2002).
\bibitem{bd4} L. A. Anchordoqui, S. P. Bergliaffa, and D. F. Torres, Phys. Rev. D {\bf 55}, 5226 (1997).
\bibitem{bd3} K. K. Nandi, B. Bhattacharjee, S. M. K. Alam, and J. Evans, Phys. Rev. D {\bf 57}, 823 (1998).
\bibitem{bd3rev} A. Bhattacharya, R. Izmailov, E. Laserra, K. K. Nandi, Class. Quant. Grav. {\bf 28}, 155009 (2011).
\bibitem{bd5} R. Shaikh and S. Kar, Phys. Rev. D {\bf 94}, 024011 (2016). 
\bibitem{bd6} A. Bhattacharya, I. Nigmatzyanov, R. Izmailov, K. K. Nandi, Class. Quant. Grav. {\bf 26} 235017 (2009).
\bibitem{bd7} A. Bhadra, K. Sarkar, D. P. Datta, K. K. Nandi, Mod. Phys. Lett. A {\bf 22}, 367 (2007);\\K. K. Nandi, I. Nigmatzyanov, R. Izmailov, N. G. Migranov, Class. Quant. Grav. {\bf 25} 165020 (2008);\\ E. F. Eiroa, M. G. Richarte, C. Simeone, Phys. Lett. A {\bf 373}, 1 (2008);\\F. S. N. Lobo, M. A. Oliveira, Phys. Rev. D {\bf 81}, 067501 (2010);\\E. F. Eiroa \& C. Simeone, Phys. Rev. D {\bf 82}, 084039 (2010);\\P. S. Letelier and A. Wang, Phys. Rev. D {\bf 48}, 631 (1993);\\F. S. Accetta, A. Chodos, Bin Shao, Nuc. Phys. B {\bf 333}, 221 (1990);\\XG. Xiao, B. J. Carr, L. Liu, Gen. Relativ. Gravit {\bf 28}, 1377 (1996);\\L. A. Anchordoqui, A. G. Grunfeld, D. F. Torres, Grav. Cosmol. {\bf 4} 287 (1998).
\bibitem{Jordan1950} P. Jordan, \lq{}\lq{}{\it Schwerkraft und Weltall: Grundlagen der theoretischen Kosmologie,}\rq{}\rq{} Friedrich Vieweg und Sohn, Braunschweig (1955);\\P. Jordan, Z. Physik {\bf 157}, 112 (1959).
\bibitem{maeda2003} Y. Fujii \& K.-ichi Maeda, \lq{}\lq{}{\it The Scalar-Tensor Theory of Gravitation,}\rq{}\rq{} United Kingdom: Cambridge University Press (2003).
\bibitem{Faraoni} V. Faraoni, \lq{}\lq{}{\it Cosmology in Scalar Tensor Gravity,}\rq{}\rq{} (Dordrecht: Kluwer Academic, 2004);\\T. Clifton, P. G. Ferreira, A. Padilla, C. Skordis, Phys. Rep. {\bf 513}, 1 (2012).
\bibitem{Singh1983} T. Singh and L. N. Rai, Gen. Relat. Gravit. {\bf 15}, 875 (1983);\\ C. Santos and R. Gregory, Annals Phys. {\bf 258}, 111 (1997);\\M. K. Mak and T. Harko, Europhys. Lett. {\bf 60}, 155 (2002);\\ Y. Bisabr, Astrophys. Space Sci. {\bf 339} 87 (2012);\\S. Chattopadhyay, A. Pasqua, M. Khurshudyan, Eur. Phys. J. C, {\bf 74}, 3080 (2014);\\M. Artymowski, Z. Lalak, M. Lewicki, JCAP 06 (2015) 031;\\ Y. Bisabr, Ann. Phys. {\bf 476}, 169962 (2025).
\bibitem{CMWill} C. M. Will, Living Rev. Relativ. {\bf 17}, 4 (2014).
\bibitem{Berto2003} B. Bertotti, L. Iess, P. Tortora, Nature {\bf 425}, 374 (2003);\\V. Faraoni, J. Cote, A. Giusti, Eur. Phys. J. C {\bf 80}, 132 (2020).
\bibitem{omega500} V. Acquaviva and L. Verde, JCAP 12 (2007) 001.
\bibitem{omegam120} F.-Q. Wu and X. Chen, Phys. Rev. D {\bf 82}, 083003 (2010).
\bibitem{nsbhgw} R. Niu, X. Zhang, B. Wang, W. Zhao, ApJ {\bf 921}, 149 (2021).
\bibitem{nsbhgw1} T. Jiang, N. Dai, Y. Gong, D. Liang, C. Zhang, JCAP 12 (2022) 023.
\bibitem{Dabrow2007} M. P. Dabrowski, T. Denkiewicz, D. Blaschke, Annalen Phys. {\bf 519}, 237 (2007).
\bibitem{BDpath} F. Hammad, D. K. Ciftci, V. Faraoni, Eur. Phys. J. Plus {\bf 134}, 480 (2019). 
\bibitem{omega32} M. Salgado, Class. Quant. Grav. {\bf 23}, 4719 (2006).
\bibitem{GrMTW} C. W. Misner, K. S. Thorne, J. A. Wheeler, \lq{}\lq{}{\it Gravitation,}\rq{}\rq{} United Kingdom: Princeton University Press (2017).
\bibitem{LOBOBDEXOT2011} N. M. Garcia, F. S. N. Lobo, Mod. Phys. Lett. A, {\bf 40} 3067 (2011).
\bibitem{polyeos} U. S. Nilsson and C. Uggla, Ann. Phys. {\bf 286}, 292 (2001).
\bibitem{polyeos1} L. Rezzolla and O. Zanotti, {\it Relativistic Hydrodynamics,} United Kingdom, Oxford (2013).
\bibitem{quadeos} T. Feroze, A. A. Siddiqui, Gen. Relativ. Gravit. {\bf 43}, 1025 (2011);\\S. D. Maharaj and P. M. Takisa, Gen. Relativ. Gravit. {\bf 44}, 1419 (2012);\\J. M. Sunzu and A. V. Mathias, Indian J. Phys. {\bf 96}, 4059 (2022).
\bibitem{polchapeos} A. K. Prasad, J. Kumar, A. Sarkar, Gen. Relativ. Gravit. {\bf 53}, 108 (2021);\\J. M. Sunzu, A. V. Mathias, Indian J. Phys. {\bf 97}, 687 (2023);\\ D. Bhattacharjee, P. K. Chattopadhyay, Eur. Phys. J. C {\bf 84}, 77 (2024).
\bibitem{vandereos} S. Capozziello, S. De Martino, M. Falanga, Phys. Lett. A {\bf 299}, 494 (2002);\\F. S. N. Lobo, Phys. Rev. D {\bf 75}, 024023 (2007);\\R. C. S. Jantsch, M. H. B. Christmann, G. M. Kremer, Int. J. Mod. Phys. D {\bf 25}, 1650031 (2016).
\bibitem{logeos} P. -H. Chavanis, Eur. Phys. J. Plus {\bf 130}, 130 (2015);\\ H. Benaoum, P.-H. Chavanis, H. Quevedo, Universe {\bf 2022}, 8, 468.
\bibitem{coscons} L. Amendola and S. Tsujikawa, \lq{}\lq{}{\it Dark Energy: Theory and Observations,}\rq{}\rq{} Cambridge University Press (2010);\\ I. B.-Dayan and U. Kumar, Eur. Phys. J. C {\bf 84}, 167 (2024);\\ L. A. Escamilla, W. Giare, E. D. Valentino, R. C. Nunes, S. Vagnozzi, JCAP 05 (2024) 091.
\bibitem{quint} V. Sahni and A. Starobinsky, Int. J. Mod. Phys. D, {\bf 9}, 373 (2000);\\ S. K. Tripathy, S. K. Pradhan, Z. Naik, D. Behera, B. Mishra, Phys. Dark Univ. {\bf 30}, 100722 (2020);\\ A. Alho, C. Uggla, J. Wainwright, Phys. Dark Univ. {\bf 39}, 101146 (2023).
\bibitem{darkphantom} A. N Baushev, J. Phys.: Conf. Ser. {\bf 203} 012055 (2010);\\K. J. Ludwick, Mod. Phys. Lett. A {\bf 32}, 28 (2017);\\ A. Bouali, I. Albarran, M. B.-Lopez, T. Ouali, Phys. Dark Univ. {\bf 26}, 100391 (2019).
\bibitem{phantwormhool} S. V. Sushkov, Phys. Rev. D {\bf 71}, 043520 (2005);\\M. Cataldo and P. Meza, Phys. Rev. D {\bf 87}, 064012 (2013);\\M. Cataldo and F. Orellana, Phys. Rev. D {\bf 96}, 064022 (2017);\\K. A. Bronnikov, Phys. Rev. D {\bf 106}, 064029 (2022);\\T. Karakasis, E. Papantonopoulos, C. Vlachos, Phys. Rev. D {\bf 105}, 024006 (2022);\\M. Sobak, Commun. Math. Phys. {\bf 405}, 232 (2024);\\ S. Pradhan, Z. Hassan, P. K. Sahoo, Phys. Dark Univ. {\bf 46}, 101620 (2024).
\bibitem{PoissonBook} E. Poisson, \lq{}\lq{}{\it A Relativist's Toolkit: The Mathematics of Black-Hole Mechanics,}\rq{}\rq{} Cambridge University Press (2004).
\bibitem{Hochberg1998} D. Hochberg and M. Visser, Phys. Rev. D {\bf 58}, 044021 (1998).
\bibitem{nonzeroredsh} P. K. F. Kuhfittig, Adv. Stud. Theor. Phys. {\bf 5}, 365 (2011);\\N. Godani, G. C. Samanta, New Astronomy, {\bf 80}, 101399 (2020).
\bibitem{catal2017} M. Cataldo, L. Liempi, P. Rodriguez, Eur. Phys. J. C {\bf 77}, 748 (2017).
\bibitem{niltonuni} M. Nilton, G. Alencar, Universe {\bf 7}, 332 (2021).
\bibitem{casschwa} A. C. L. Santos, C. R. Muniz, and L. T. Oliveira, Int. J. Mod. Phys. D {\bf 30}, 2150032 (2021).
\bibitem{AppellHyper} L. J. Slater, \lq{}\lq{}{\it Generalized Hypergeometric Functions,}\rq{}\rq{} Cambridge University Press (1966);\\I. S. Gradshteyn \& I. M. Ryzhik, \lq{}\lq{}{\it Table of Integrals, Series, and Products,}\rq{}\rq{} United States: Academic Press (2014).
\bibitem{singgrbd} S. W. Hawking \& G. F. R. Ellis, \lq{}\lq{}{\it The Large Scale Structure of Space-Time,}\rq{}\rq{} Cambridge University Press (1975).
\bibitem{singgrbd1} S. W. Hawking, Comm. Math. Phys. {\bf 25}, 167 (1972);\\T. P. Sotiriou \& V. Faraoni, Phys. Rev. Lett. {\bf 108}, 081103 (2012);\\ M. Campanelli, \& C. O. Lousto, Int. J. Mod. Phys. D {\bf 02} 451-462 (1993);\\N. Bedjaoui, P. G. LeFloch, J. M. M.-Garcia, J. Novak, Class. Quant. Grav. {\bf 27}, 245010 (2010);\\D. A. Tretyakova, B. N. Latosh, S. O. Alexeyev, Class. Quant. Grav. {\bf 32}, 185002 (2015).
\bibitem{singuissues} J. Earman, \lq{}\lq{}{\it Bangs, Crunches, Whimpers, and Shrieks Singularities and Acausalities in Relativistic Spacetimes,}\rq{}\rq{} United Kingdom: Oxford University Press (1995);\\J. Earman, Found. Phys. {\bf 26}, 623 (1996);\\G. E. Romero, Found. Sci. {\bf 18}, 297 (2013);\\ N. Sfetcu, \lq{}\lq{}{}{\it The Singularities As Ontological Limits of the General Relativity,}\rq{}\rq{} N.p., Lulu.com (2019);\\ F. Azhar, M. H. Namjoo, arXiv:2101.10887 [physics.hist-ph].
\bibitem{GRSINDOM} A. D. Rendall, arXiv:gr-qc/0503112.
\bibitem{GRSINDOM1} J. M. M. Senovilla, Gen. Relativ. Gravit. {\bf 30}, 701 (1998).
\bibitem{Ellis1977} G. F. R. Ellis, B. G. Schmidt, Gen. Relat. Gravit. {\bf 8}, 915 (1977).
\bibitem{sing1979} C. B. Collins \& G. F. R. Ellis, Phys. Rep. {\bf 56}, 67 (1979).
\bibitem{KRSING} P. S. Joshi, \lq{}\lq{}{\it Gravitational Collapse and Spacetime Singularities,}\rq{}\rq{} United Kingdom: Cambridge University Press (2012).
\bibitem{Racz2023} I. Racz, Gen. Relativ. Gravit. {\bf 55}, 3 (2023).
\bibitem{valerio2016} V. Faraoni, F. Hammad, S. D. Belknap-Keet, Phys. Rev. D {\bf 94}, 104019 (2016).
\bibitem{SPR1393} A. Ashtekar \& V. Petkov, (eds.) \lq{}\lq{}{\it Springer Handbook of Spacetime,}\rq{}\rq{} Springer, Berlin, (2014).
\bibitem{Liu2024} S. H. Mazharimousavi and M. Halilsoy, Mod. Phys. Lett. A {\bf 31}, 1650192 (2016);\\N. S. Kavya, V. Venkatesha, G. Mustafa, P. K. Sahoo, S. V. D. Rashmi, Chinese J. Phys. {\bf 84}, 1 (2023);\\ V. D. Falco and S. Capozziello, Phys. Rev. D, {\bf 108}, 104030 (2023);\\J. Lu, S. Yang, Yan Liu, Y. Zhang, Yu Liu, Eur. Phys. J. Plus, {\bf 139}, 274 (2024);\\ T. Tangphati, A. Banerjee, A. Pradhan, Int. J. Geom. Meth. Mod. Phys. {\bf 21}, 2450109 (2024).
\bibitem{Hornworm} O. A. Evseev and O. I. Melichev, Phys. Rev. D {\bf 97}, 124040 (2018);\\S. Mironov, V. Rubakov, V. Volkova, Class. Quantum Grav. {\bf 36}, 135008 (2019);\\R. Korolev, F. S. N. Lobo, S. V. Sushkov, Phys. Rev. D {\bf 101}, 124057 (2020);\\A. Bakopoulos, C. Charmousis, P. Kanti, JCAP 05 (2022) 022;\\A. Bakopoulos, N. Chatzifotis, C. Erices, E. Papantonopoulos, JCAP 11 (2023) 055.
\bibitem{semibdquant} L. J. Garay and J. G.-Bellido, Nucl. Phys. B {\bf 400}, 416 (1993);\\ H.-H. Yang, and Y. Z. Zhang, Phys. Lett. A {\bf 212}, 39 (1996).
\bibitem{bd1loop} A. A. Tseytlin, Phys. Lett. B {\bf 176}, 92 (1986);\\ C. G. Callan, C. Lovelace, C. R. Nappi, S. A. Yost, Nucl. Phys. B {\bf 288}, 525 (1987);\\D. Blaschke and M. P. Dabrowski, Entropy {\bf 14}, 1978 (2012);\\Z. Haba, hep-th/0205130 [hep-th].
\bibitem{quantumworm} M. Visser, Phys. Rev. D {\bf 43}, 402 (1991);\\I. Redmount and W.-M. Suen, Phys. Rev. D {\bf 49}, 5199 (1994);\\L. Susskind and Y. Zhao, Phys. Rev. D {\bf 98}, 046016 (2018);\\ F. Tamburini and I. Licata, Entropy {\bf 2020}, 22, 3.
\end{thebibliography}
\end{document}